# Discrete Quantum Electrodynamics

Charles Francis

**Abstract:**

The purpose of this paper is to construct a quantum field theory suitable for describing quantum electrodynamics and Yang-Mills theory in a form which satisfies the conditions of the Millennium prize offered by the Clay Mathematics Institute as described by Jaffe and Witten [12], by showing that it satisfies '*axioms at least as strong as those cited by*' Wightman [18] and by Osterwalder and Schrader [14], and by observing that this form of field theory has no mass gap. The definitions provide a model for relativistic quantum mechanics which supports a form of relativistic quantum field theory, but which does not depend on the second quantisation of a "matter wave". Continuous laws of wave mechanics are found in model of discrete particle interactions which does not involve waves, or the quantisation of interacting fields. Newton's first law and conservation of momentum are established from the principle of homogeneity. Maxwell's equations are derived from the simple interaction in which a Dirac particle emits or absorbs a photon, showing that the renormalised mass and coupling constant are equal to their bare values. Feynman rules are calculated for the discrete theory and give the predictions of the standard renormalised theory. Quark confining interactions are described for qed and for an adaptation of Yang-Mills theory.

Charles Francis
Clef Digital Systems Ltd.
Lluest, Neuaddlwyd
Lampeter
Ceredigion
SA48 7RG

# Discrete Quantum Electrodynamics

## 1  Background

It has long been known that there is no finite unitary representation of the Lorentz group, and this paper seeks to resolve the problems in constructing field operators on covariant Hilbert space by defining equivalence classes of operators on an infinite family of finite dimensional Hilbert spaces. Lorentz transformation will invoke a different Hilbert space on which the transformed field operator is unambiguously defined. A lattice is used to define the basis for each Hilbert space, but there are important differences between discrete quantum field theory as described here and, e.g. lattice quantum field theory as developed by Wilson and others and described by e.g. Montvay & Münster [13]. The physical structure to which this mathematical model relates is also described using an "information space" interpretation in which reality exists but quantum mechanics describes the information we have about it. The reader should distinguish interpretation from the mathematical construction based on the given definitions which is the subject of this paper, but interpretation is necessary because the mathematical structure applies to each observer's information space while it could not apply to ontological space-time, and to avoid confusion with interpretational remarks found in other treatments which cannot be applied here. The interpretation is orthodox, being based on Heisenberg's discussion of the Copenhagen interpretation in [9] and the Dirac-Von Neumann approach to it [10][2]. It does not use Bohr's complementary, Jordan's second quantisation of 'matter fields', nor have a dependency on a classical Lagrangian or action principle. Commutation relations are found from the definition of momentum states as linear combinations of position states, not simply imposed by canonical quantisation.

As distinct from lattice qed, discrete qed uses Minkowsky rather than Euclidean co-ordinates, and uses a bounded momentum space with an automatic cut-off. An energy cut-off follows from the mass shell condition, but there is no cut-off in the off mass shell energy which appears in the perturbation expansion, since this is an abstract parameter in a contour integral (section 18). A discrete model cannot be manifestly covariant. Manifest covariance is not necessary since the lattice is not physical and is observer dependent. It is required that the laws of physics are the same in all reference frames, and covariance is redefined to allow an *invariant* hermitian product on a space of *covariant* functions. Points in one observer's information space do not transform to those in another's and operators for position in different reference frames do not commute. Classical vector observables exist on a scale where the lattice appears as a continuum. This scale appears many orders of magnitude smaller than experimental errors. Manifest covariance and the renormalised formulae of the standard model are recovered on letting lattice size go to infinity and lattice spacing go to zero. The Landau pole appears in this limit but only after reordering series where we do not have absolute convergence and does not indicate anything more serious than the breakdown of an iterative solution to a non-perturbative equation. In fact the Landau pole does not appear if lattice spacing is bounded below by a fundamental interval of proper time in the interactions between particles, and it is suggested that this is in fact the case in a subsequent paper [7], where a minimum proper time between interactions is related to unification with gravity.

Using finite dimensional space the definition of linear field operators is unproblematic, as they are operators, not operator valued distributions. The field operators constructed here will be used to describe interactions between particles using the interaction picture. Although formally similar to quantised free fields, they will be used to describe the potential for the creation or annihilation of a particle in an interac-tion. This cannot be reconciled to interpretational statements which are sometimes made about models which arise from the quantisation of classical fields, such as "*The free field describes particles which do not interact*" (Glimm & Jaffe p 100 [8]); it will be appreciated that while the formulae of the present construction are extremely like the formulae which arise in such models and give the same physical predictions in the appropriate limits, the meaning of these formulae is quite different.

Although apparently innocuous, perhaps the most significant mathematical difference between this and other formulations of field theory is that time is a parameter, as in non-relativistic quantum mechanics, and each Hilbert space $\mathscr{H}(t)$ is formulated for synchronous states at time $t$. The U-matrix is strictly a map $U(t_1, t_2): \mathscr{H}(t_1) \to \mathscr{H}(t_2)$ where $t_1 \neq t_2$, so that unitarity does not apply (conservation of probability is required). Homomorphically identifying $(\forall t) \mathscr{H}(t) = \mathscr{H}$ introduces what has perhaps been the central problem in the construction of quantum field theory in 4 dimensions, namely the indefinability of the equal point multiplication between the field operators (which is definable but would invalidate the limit as lattice spacing goes to zero). This is resolved pragmatically by normal ordering all equal point multiplications of field operators. This non-linear condition is motivated physically by saying that a particle created at $x_0^n$ cannot interact again at the same instant. This has an important effect upon renormalisation and gives physical motivation to the method of Epstein & Glaser [17][5] in which $\theta(t_1 - t) - \theta(t - t_1)$ is replaced with a continuous switching function which is zero at $t = t_1$, removing loop divergences. The analysis of the origin of the ultraviolet divergence is, for practical purposes, that given by Scharf [17], namely the incorrect use of Wick's theorem. The difference between this treatment and Scharf is that here the limiting procedure uses a discrete sum whereas Epstein and Glaser use a continuous switching function, and while Scharf says (p163) "the switching on and off the interaction is unphysical", here the switching off and on of the interaction at $x_0^n = x_0^i$ is a physical constraint meaning that only one interaction takes place for each particle in any instant, i.e. that the interaction operator at time $t$ cannot act on the result of itself. In practice this is done by normal ordering all equal time products of fields. Thus an annihilation operator at time $t$ does not act on a creation operator at the same time. As with the method of Epstein & Glaser, this leads directly to a perturbation expansion in which the terms are finite and similar in form to the standard "renormalised" expansion.

Thus we will state that the correct description of physical processes in qft uses an interaction Hamiltonian $H(x)$ such that $H^2(x) = 0$. $H$ is not a linear operator so we do not have $H(x) = 0$. In other respects $H$ is much as in ordinary qed. Unitarity does not apply to non-linear operators, and it will be shown that $H$ preserves the norm as required by conservation of probability. Linearity of time evolution is required to avoid a dependency on histories; the interactions of a particle created yesterday should be no different from those of one created the day before. Despite the apparently radical nature of non-linearity it will be seen that the time evolution proposed here is linear for states created in the past, and only distinguishes the interaction of a particle created "now" from those of particles created "not now".

It will be clear from the use of families of finite dimensional Hilbert spaces that the present construction does not obey the Wightman axioms, which specify a single covariant Hilbert space. It will be clear once transformations between observer dependent Hilbert spaces have been specified, and field operators have been defined on such Hilbert spaces, that the model obeys *axioms at least as strong as those cited by* Wightman [18] and by Osterwalder and Schrader [14]. It is immediate that the algebras of operators defined by 15.18 for the photon, and by 16.1 together with 16.5 and 16.8 for the Dirac field obey the Haag-Kastler axioms given in [8] after redefining the Lorentz group as specified in section 8. It will be seen that bare particle masses are a parameter of this model, independent of lattice spacing, and they are not changed by renormalisation. The existence of a mass gap is equivalent to the statement that the model has no zero mass particles. The model allows a zero mass neutrino (even if there is not one in nature), and since photons, and other intermediate vector bosons such as the gluon, are here treated as particles on an equal footing with fermions, it is obvious that this model has no mass gap. Indeed there will always be eigenvalues of the Hamiltonian containing an indefinite number of photons at an energy threshold below that for particle creation. Naturally if a massive particle is present, the spectrum of the Hamiltonian will always be above that mass.

## 2   Information Space

It is important to distinguish the definitional properties of information space from postulates about the behaviour of matter. We make no assumption of an ontological background in which matter is placed and define a reference frame as the set of potential results of measurement of position. In this view geometry is simply and literally world (*geo-*) measurement (*-metry*); to understand geometry we must study how observers measure space-time co-ordinates. Each observer has a clock, which is, without loss of generality, the origin of his co-ordinate system and measures proper time for that observer. Each event is given a co-ordinate by measuring the time taken for light to travel to and fro the event. Now suppose there is some fundamental minimum unit of time. Then the measured co-ordinates are integer multiples of that unit and the co-ordinate system is a lattice consisting of measured events and potential measured events in discrete multiples of that unit, bounded by the time period in which the observer carries out measurements. Thus each observer's reference frame is defined as a lattice determined by the finite resolution with which it is possible to measure time; it is part of information space of a particular observer, not prior ontology. For simplicity we will use a cubical lattice, although any lattice can be used as appropriate to the measuring apparatus and chosen co-ordinate system of an individual observer.

The notion that space-time appears not as an invariant background but as an observer dependent set of potential measurement results, is in strictly accordance with the orthodox interpretation of quantum mechanics. In Dirac's words "*In the general case we cannot speak of an observable having a value for a particular state, but we can .... speak of the probability of its having a specified value for the state, meaning the probability of this specified value being obtained when one makes a measurement of the observable*" [3]. When this statement is applied to the position observable, it follows that position exists only in measurement of position, and not as part of background geometry.

The measurement of time and position is sufficient for the study of many (it has been suggested all) other physical quantities; for example a classical measurement of velocity may be reduced to a time trial over a measured distance, and a typical measurement of momentum of a particle involves plotting its path in a bubble chamber. In practice there is also a bound on the magnitude of the result, and we take the results of measurement of position to be in a finite region $\mathrm{N} \subset \mathbb{N}^3$, in units of lattice spacing, $\chi$. Without loss of generality define

**Definition:** *The coordinate system is* $\mathrm{N} = (-\nu, \nu] \otimes (-\nu, \nu] \otimes (-\nu, \nu] \subset \mathbb{N}^3$ *for some* $\nu \in \mathbb{N}$, *where* $(-\nu, \nu] = \{x \in \mathbb{N} | -\nu < x \leq \nu\}$.

There is no significance in the bound, ν, of a given co-ordinate system N. If matter goes outside of N it is merely moving out of a co-ordinate system, not out of the universe. Generally it is possible to describe its motion in another co-ordinate system with another origin. Even if we do not intend to take the limit $\nu \to \infty$, N may chosen large enough to say with certainty up to the limit of experimental accuracy, that it contains any particle under study for the duration of the experiment. We ignore matter outside of N and impose conservation of probability as usual.

**Definition:** *Let $\mathcal{N}$ be the set of finite discrete coordinate systems of sufficient size to include all particles under study which can be defined by an observer by means of physical measurement. $\mathcal{N}$ includes curvilinear as well as rectangular coordinates.*

The intention is to compose a manifold out of the collection of such co-ordinate systems. In general this will require non-Euclidean metric [7], but for the purpose of this paper, co-ordinate systems in $\mathcal{N}$ will be defined using a flat metric, and will be regarded as defining a tangent space. We will not here study the connection between coordinate systems with different origins, but merely note the intention that this model be compatible with any Lorentzian manifold.

**Definition:** *For any point $x \in N$, $|x\rangle$ is the ket corresponding to a measurement of position with result x. $|x\rangle$ is called a position ket.*

**Definition:** *Let $|\rangle$ be the empty ket or the vacuum state, a name for a state of no particles.*

In the absence of information, we cannot describe the actual configuration of particles; kets are names or labels for states, not descriptions of matter. The significance is that the principle of superposition will be introduced as a definitional truism in a naming system, not as a physical assumption. Although kets are not states, but merely names for states, we loosely refer to kets as states in keeping with common practice when no ambiguity arises. In a typical measurement in quantum mechanics we study a particle in near isolation. The suggestion is that there are too few ontological relationships to generate the property of position. Then position does not exist prior to the measurement, and the measurement itself is responsible for introducing interactions which generate position. In this case, prior to the measurement, the state of the system is not labelled by a position ket, and we define labels containing information about other states – namely the information about what would happen in a measurement.

In scientific measurement we set up many repetitions of a system, and record the frequency of each result. Probability is interpreted here simply as a prediction of a frequency distribution, so a mathematical model must associate a probability with each possible result. The probability of a given result can be used to attach a label to the state in the following way:

**Definition:** *Let $\mathbb{H}_0 = \{|x\rangle | x \in N\} \cup \{|\rangle\}$*

**Definition:** *Construct a vector space, $\mathbb{H}$, over $\mathbb{C}$, with basis $\mathbb{H}_0$. The members of $\mathbb{H}$ are called kets.*

This is trivial because $\mathbb{H}_0$ is finite. $\mathbb{H}$ has dimension $8\nu^3 + 1$.

**Definition:** $\forall |f\rangle, |g\rangle \in \mathbb{H}$, *the braket $\langle g | f \rangle$ is the hermitian form on $\mathbb{H}$ defined by its action $\mathbb{H}_0$*

$$\forall x, y \in N, \langle x|y \rangle = \delta_{xy}, \langle | \rangle = 1, \langle | x \rangle = \langle x | \rangle = 0 \qquad 2.1$$

**Definition:** *The position function of the ket $|f\rangle \in \mathbb{H}$ is the mapping $N \to \mathbb{C}$ defined by*

$$\forall x \in N, x \to \langle x|f \rangle$$

Later the position function will be identified with the restriction of the wave function to $\mathbb{N}$, but we use the term position function, because it is discrete and because a wave is not assumed. We observe that

$$P(x/f) = \frac{|\langle f | x \rangle|^2}{\langle f | f \rangle} \qquad 2.2$$

has the properties of a probability distribution, and we associate the ket $|f\rangle \in \mathbb{H}$ with any state such that $\forall x \in \mathbb{N}$ the probability that a measurement of position has result $x$ is given by $P(x|f)$.

In this interpretation the position function (and more generally the scalar product) can be identified with a proposition in a many valued logic [16]. The value of the position function is the (complex) truth value of the proposition. Classical logic applies to sets of statements about the real world which are definitely true or definitely false. For example, when we make a statement

$$\mathscr{P}(x) = \text{The position of a particle is } x \qquad 2.3$$

we tend to assume that it is definitely true, or definitely false. If it is actually the case that $\mathscr{P}(x)$ is definitely true or false, then classical logic and classical mechanics apply. But in quantum mechanics it is not, in general, possible to say that a particle has a definite position when position is not measured, and we can only consider sets of propositions to describe hypothetical measurement results

$$\mathscr{T}(x) = \text{If a measurement of position were to be done the result would be } x \qquad 2.4$$

The only empirical states are those which are measured, and only for measured states can we make definite statements with in the form of $\mathscr{P}(x)$ and having truth values from the set $\{0, 1\}$. But we can categorise other states according to the likelyhood of what might happen in measurement by using the structure of Hilbert space and a probability interpretation based on the inner product. Thus we construct new propositions by identifying the vector sum $a|f\rangle + b|g\rangle$ with weighted logical OR between propositions. We say that $\langle x|f\rangle$ is the truth value of the proposition $\mathscr{T}(x)$. Multiplication by a scalar only has meaning as a weighting between alternatives and we have that $\forall |f\rangle \in \mathbb{H}, \forall \lambda \in \mathbb{C}$ such that $\lambda \neq 0$, $\lambda|f\rangle$ labels the same state (or set of states) as $|f\rangle$. Probabilities defined from 2.2 are invariant under local $U(1)$ gauge symmetry and interactions must preserve this symmetry since it describes redundant information appearing in the mathematical model but without physical meaning. The information space interpretation inverts the measurement problem. Collapse just represents a change in information due to a new measurement. But Schrödinger's equation does require explanation – interference patterns are real. It will be derived in 9.1. It is routine to prove that

$$\langle g|f\rangle = \sum_{x \in \mathbb{N}} \langle g|x\rangle\langle x|f\rangle \qquad 2.5$$

$$\sum_{x \in \mathbb{N}} |x\rangle\langle x| = 1 \qquad 2.6$$

So long as we recognise that these are just rules, a calculational device which says nothing about metaphysics there is no inconsistency, ambiguity or other problem with the use of Hilbert space to categorise unmeasured states. The wave property of superposition comes from the logic rather than the metaphysic. We can make statements in the form of $\mathscr{T}(x)$ and interpret the probability amplitude for such statements as a truth value, but we cannot in general make statements in the form $\mathscr{P}(x)$. The interpretation begs the question *what is wrong with* $\mathscr{P}(x)$? Is there no such thing as numerical position except in certain states, or is there no such thing as a particle? If in fact there is no background continuum then there is no reason to reject the notion that the universe is composed of particles.

## 3  Momentum Space

In this treatment momentum is defined mathematically, not empirically. The correspondence with a measurable physical quantity is to be made by demonstrating conservation of momentum (section 14). Although the formulae derived below are largely standard it is necessary to show them here to distinguish definitional properties of a mathematical structure from physical postulates, because canonical quantisation is not assumed, and because this treatment uses finite dimensional Hilbert space which has useful properties which do not hold in infinite dimensional space. In particular it is worth observing that representations of the Dirac delta function on momentum space are ordinary functions (a distribution theoretic treatment is recovered by allowing lattice spacing to tend to zero *after* calculation of probabilities).

**Definition:** *Momentum space is the 3-torus* $M = [-\pi, \pi] \otimes [-\pi, \pi] \otimes [-\pi, \pi] \subset \mathbb{R}^3$. *Elements of momentum space are called momenta.*

**Definition:** *For each value of momentum* $p \in M$, *define a ket* $|p\rangle$, *known as a plane wave state, by the position function*

$$\langle x | p \rangle = \left(\frac{1}{2\pi}\right)^{\frac{3}{2}} e^{ix \cdot p} \qquad 3.1$$

The unit of momentum is 1/length. If lattice spacing is $\chi$ in conventional units, then there is a bound of $\pi/\chi$ on the magnitude of each component of momentum. In general this is a bound on the magnitude of momentum which can be measured with apparatus at any given resolution, but later we will interpret $\chi$ as a fundamental length determined by the interactions of matter. Then $\chi$ is a fundamental physical constant which is invariant in all reference frames, and $\pi/\chi$ is an absolute physical bound on momentum. The implications for transformations of a bound on momentum with be considered in more detail in section 8. We will use units in which $\chi = 1$. The expansion of $|p\rangle$ in the basis $\mathbb{H}_0$ is calculated by using the resolution of unity, 2.6

$$|p\rangle = \sum_{x \in N} |x\rangle\langle x|p\rangle = \left(\frac{1}{2\pi}\right)^{\frac{3}{2}} \sum_{x \in N} e^{ix \cdot p} |x\rangle \qquad 3.2$$

**Definition:** *For each ket* $|f\rangle$ *define the momentum space function as the Fourier transform,* $F: M \to \mathbb{C}$,

$$F(p) = \langle p | f \rangle \qquad 3.3$$

By 2.6, $F$ can be expanded as a trigonometric polynomial

$$F(p) = \sum_{x \in N} \langle p | x \rangle \langle x | f \rangle = \left(\frac{1}{2\pi}\right)^{\frac{3}{2}} \sum_{x \in N} \langle x | f \rangle e^{-ix \cdot p} \qquad 3.4$$

by 3.1 and 3.3. By 3.4 and 3.1

$$\left(\frac{1}{2\pi}\right)^{\frac{3}{2}} \int_M d^3p \, F(p) e^{ix \cdot p} = \frac{1}{8\pi^3} \int_M d^3p \sum_{y \in N} \langle x | f \rangle e^{-iy \cdot p} e^{ix \cdot p} = \langle x | f \rangle \qquad 3.5$$

Rewriting 3.5 in the notations of 3.3 and 3.1

$$\langle x | f \rangle = \int_M d^3p \langle x | p \rangle \langle p | f \rangle \qquad 3.6$$

For any integrable $F':\mathrm{M}\to\mathbb{C}$ there is a unique position function given by

$$\langle x|f\rangle = \left(\frac{1}{2\pi}\right)^{\frac{3}{2}} \int_\mathrm{M} d^3p\, F'(p) e^{ix\cdot p} \qquad 3.7$$

In general 3.7 is not invertible, but 3.4 picks out a unique invertible member of the equivalence class of functions $F':\mathrm{M}\to\mathbb{C}$ with the same position function.

**Definition:** *Members of this class are called representations of the momentum space wave function. 3.4 is the analytic momentum space function in a given reference frame.*

The cardinality of the plane wave states is greater than the cardinality of $\mathbb{H}_0$, so $\{|p\rangle\,|\,p\in\mathrm{M}\}$ is not a basis. A convenient basis in momentum space is

$$\mathbb{H}_\mathrm{M} = \left\{|p\rangle\,|\,p\in\mathrm{M} \text{ and for } i = 1, 2, 3 \ \frac{\nu p_i}{\pi}\in\mathbb{Z}\right\}$$

Then it is natural to define another invertible representation of the momentum space wave function:

**Definition:** *The discrete momentum space wave function is defined by the DFT (Discrete Fourier Transform)*

$$F_D(p) = F(p^b) \text{ where } p^b \in \mathbb{H}_\mathrm{M} \text{ is such that for } i = 1, 2, 3 \ \ 0\le p_i - p_i^b < \pi/\nu \qquad 3.8$$

*Proof:* 3.8 is a representation because 3.7 reduces to

$$\left(\frac{1}{2\pi}\right)^{\frac{3}{2}} \int_\mathrm{M} d^3p\, F_D(p) e^{ix\cdot p} = \left(\frac{1}{2\pi}\right)^{\frac{3}{2}} \sum_{|p\rangle\in\mathbb{H}_\mathrm{M}} \langle p|f\rangle e^{ix\cdot p}$$

$$= \left(\frac{1}{2\pi}\right)^3 \sum_{|p\rangle\in\mathbb{H}_\mathrm{M}} \sum_{x'\in\mathrm{N}} \langle x'|f\rangle e^{-ix'\cdot p} e^{ix\cdot p}$$

$$= \langle x|f\rangle$$

The use of the DFT extends the definitions of M and N in a non-physical manner so that both the coordinate space and momentum space wave functions are periodic. This has no physical meaning. Since energy is conserved in physical measurement (section 18), momentum is bounded by the total energy of a system. The probability of finding a momentum above this bound is zero. In general the analytic representation does not have support with such a bound, but exhibits Gibb's ripples. The analytic representation $\langle p|f\rangle$ is not directly related to probability, but the discrete representation which uses a basis in momentum space is a probability amplitude, and we assume that for physically realisable states the support of the discrete representation of $\langle p|f\rangle$ has a suitable bound. The importance of this bound is further discussed in section 8. Because the model is defined in position space with the hermitian product given by 2.5 all predictions are identical for each equivalent momentum space wave function. For example, the components of the experimental momentum operator are given by

$$\text{for } i = 1, 2, 3\,, \ P_i = \frac{-i}{2}\sum_{x\in\mathrm{N}} |x\rangle[\langle x+\mathbf{1}_i| - \langle x-\mathbf{1}_i|]$$

Then

$$P_i|p\rangle = \frac{-i}{2}\sum_{x\in\mathrm{N}} |x\rangle[\langle x+\mathbf{1}_i| - \langle x-\mathbf{1}_i|]|p\rangle = \sum_{x\in\mathrm{N}} |x\rangle\langle x|p\rangle \sin p_i = \sin p_i|p\rangle \qquad 3.9$$

So the eigenvalue of momentum is $\sin p \approx p$ for $p$ much less than $\pi/\chi$. There is some indication that magnitude of the discrete unit of time for a particular particle is twice the Schwarzschild radius $4Gm/c^3$, where $m$ is the mass of the particle and $G$ is the gravitation constant [7]. Then an electron with a difference of 0.1% between $p$ and $\sin(p)$ would have an energy of $0.055\pi/\chi = 1.38 \times 10^{52}$ eV, which may be thought unrealistic.

By 2.6 and by 3.6 $\forall |f\rangle, |g\rangle \in \mathbb{H}$

$$\langle g|f\rangle = \sum_{x \in \mathbb{N}} \langle g|x\rangle\langle x|f\rangle = \int_M d^3p \sum_{x \in \mathbb{N}} \langle g|x\rangle\langle x|p\rangle\langle p|f\rangle = \int_M d^3p \langle g|p\rangle\langle p|f\rangle \qquad 3.10$$

3.10 is true for all $|f\rangle$ and $|g\rangle$, so we can define a second form of the resolution of unity

$$\int_M d^3p |p\rangle\langle p| = 1 \qquad 3.11$$

It follows immediately that

$$\langle q|f\rangle = \int_M d^3p \langle q|p\rangle\langle p|f\rangle$$

So $\langle p|q\rangle$ has the effect of a Dirac delta function on the test space of momentum space functions.

**Definition:** *The Dirac delta function is*

$$\delta: \mathscr{T}^3 \to \mathbb{C} \qquad \delta(p-q) = \langle q|p\rangle$$

Explicitly, calculating $\langle q|p\rangle$ directly from 3.2

$$\delta(p-q) = \frac{1}{8\pi^3} \sum_{x \in \mathbb{N}} e^{ix \cdot (p-q)} \qquad 3.12$$

## 4  Multiparticle States

The rigorous construction of field operators requires the formal definition of multiparticle states and of creation and annihilation operators. In discrete qed these are operators not operator valued distributions, and we describe their definition and properties in some detail. Commutation relations will be demonstrated from the definitions, not imposed canonically. We will define notation in which bras and kets are operators as well as states. Just as Dirac notation is simpler and more elegant than wave functions, we will see that ket notation for operators is very powerful and leads to an elegant clarification of field theory.

**Definition:** *For $n \in \mathbb{N}, n \geq 0$  $\mathbb{H}^n = \bigotimes_n \mathbb{H}$ (the external direct product) is the space of kets for states of particles of the same type.*

Since $|\rangle \in \mathbb{H}_0$  $\mathbb{H}^n$ contains states of an indefinite number of particles  A one particle state cannot be a no particle state, so by the definition of the braket as a measure of uncertainty

$$\forall |f\rangle \in \mathbb{H} \quad \langle |f\rangle = 0 \qquad 4.1$$

It follows that the inner product between states of different numbers of particles is always zero. For $|f\rangle = (|f_1\rangle, ..., |f_n\rangle), |g\rangle = (|g_1\rangle, ..., |g_n\rangle) \in \mathbb{H}^n$ the braket is given by

$$\langle f|g\rangle = \langle |f_1\rangle, ..., |f_n\rangle | |g_1\rangle, ..., |g_n\rangle \rangle = \prod_{i=1}^n \langle f_i|g_i\rangle \qquad 4.2$$

as is required by the probability interpretation, 2.2, if each of the particles is independent. Let $N \in \mathbb{N}$ be larger than the number of particles in the universe. The space of all particles of a given type $\gamma$ is $\mathbb{H}^N$ The statement that we can take a value of $N$ greater than any given value is the definition of an infinite sequence, so, in effect the space of all particles of the same type is $\mathbb{H}^\infty$ Let $\gamma$ be an index running over each type of particle. The space of states of particles of type $\gamma$ is

$$\mathbb{H}^N = \bigoplus_{n=0}^{N} \mathbb{H}^n \qquad 4.3$$

Until the treatment of interactions particles of different types $\gamma$ will be ignored.

## 5 Creation Operators

The creation of a particle in an interaction is described by the action of a creation operator. Creation operators incorporate the idea that particles of the same type are identical, so that when a particle is created it is impossible to distinguish it from any existing particle of the same type. There is some advantage in using creation and annihilation operators to (anti)symmetrise states and at the same time to generate physical space, since this ties in with the idea that physical states are created in interactions which are themselves described as combinations of creation and annihilation operators, and because this will simplify the treatment of quark parastatistics in section 19. Creation operators are defined by their action on the basis states $\mathbb{H}_0$. The definition removes arbitrary phase and normalises the two particle state to coincide with 4.2.

**Definition:** $\forall |x\rangle \in \mathbb{H}_0$ *the creation operator* $|x\rangle : \mathbb{H}^1 \to \mathbb{H}^2$ *is defined by* $\forall |f\rangle \in \mathbb{H}^1$

$$|x\rangle : |f\rangle \to |x\rangle |f\rangle = |x;y\rangle = \frac{1}{\sqrt{2}}[(|x\rangle, |f\rangle) + \kappa(|f\rangle, |x\rangle)] \qquad 5.1$$

where $\kappa \in \mathbb{C}$ is to be determined.

**Definition:** *The bra corresponding to* $|x;y\rangle$ *is designated by* $\langle x;y|$.

A bra is here simply an alternate notation for a ket. It will be redefined as an operator in the next section. Now, by 4.2 and 5.1 $\forall x, y \in \mathbb{N}$

$$\langle x;y|x;y\rangle = \frac{1}{2}[\langle x|x\rangle\langle y|y\rangle + \kappa^2 \langle x|x\rangle\langle y|y\rangle + 2\kappa\langle x|y\rangle\langle y|x\rangle]$$

$$= \frac{1}{2}[(1+\kappa^2) + 2\kappa\delta_{xy}^2] \qquad 5.2$$

The order in which particles are created can make no difference to the state, so

$$\exists \lambda \in \mathbb{C} \text{ such that } |x;y\rangle = \lambda|y;x\rangle \qquad 5.3$$

Thus, by direct application of 4.2 and 5.1

$$\langle x;y|x;y\rangle = \lambda\langle x;y|y;x\rangle = \frac{1}{2}\lambda[\kappa\langle x|x\rangle\langle y|y\rangle + \kappa\langle x|x\rangle\langle y|y\rangle + (1+\kappa^2)\langle x|y\rangle\langle y|x\rangle]$$

$$= \frac{1}{2}\lambda[2\kappa + (1+\kappa^2)\delta_{xy}^2] \qquad 5.4$$

Comparison of 5.2 with 5.4 gives

$$1 + \kappa^2 = 2\lambda\kappa \text{ and } \lambda(1+\kappa^2) = 2\kappa \qquad 5.5$$

Hence $\lambda^2 = 1$. $\lambda = \pm 1$. Substituting into 5.5;

if $\lambda = -1$, then $1 + \kappa^2 = -2\kappa$ so $\kappa = -1$;

if $\lambda = 1$, then $1 + \kappa^2 = 2\kappa$ so $\kappa = 1$

**Definition:** *Bosons are particles for which* $\kappa = 1$, *so that* $\forall |x\rangle \in \mathbb{H}_0$ *the creation operators* $|x\rangle$ *obey*

$$\forall y \in N \quad |x;y\rangle = \frac{1}{\sqrt{2}}[(|x\rangle, |y\rangle) + (|y\rangle, |x\rangle)] = |y;x\rangle \qquad 5.6$$

**Definition:** *Fermions are particles for which* $\kappa = -1$, *so that* $\forall |x\rangle \in \mathbb{H}_0$ *the creation operators obey*

$$\forall y \in N \quad |x;y\rangle = \frac{1}{\sqrt{2}}[(|x\rangle, |y\rangle) - (|y\rangle, |x\rangle)] = -|y;x\rangle \qquad 5.7$$

The use of the ket notation for creation operators is justified by the homomorphism defined by

$$|x\rangle|\rangle = \frac{1}{\sqrt{2}}[(|x\rangle, |\rangle) + \kappa(|\rangle, |x\rangle)] \qquad 5.8$$

It is straightforward to check that this is a homomorphism with the scalar product defined by 4.2. In general the creation operator is defined by linearity

$$\forall |f\rangle \in \mathbb{H} \quad |f\rangle : \mathbb{H}^1 \to \mathbb{H}^2, \ |f\rangle = \sum_{x \in N} \langle x|f\rangle |x\rangle \qquad 5.9$$

It follows immediately that $\forall |f\rangle, |g\rangle \in \mathbb{H}$

$$|f;g\rangle = |f\rangle|g\rangle$$

$$= \sum_{x \in N} \langle x|f\rangle |x\rangle \sum_{y \in N} \langle y|g\rangle |y\rangle$$

$$= \sum_{x,y \in N} \langle x|f\rangle\langle y|g\rangle |x;y\rangle \qquad 5.10$$

Using 5.6 and 5.7 gives

$$\forall \text{ Boson } |f\rangle, |g\rangle \in \mathbb{H}^1 \quad |f;g\rangle = |g;f\rangle \qquad 5.11$$

$$\forall \text{ Fermion } |f\rangle, |g\rangle \in \mathbb{H}^1 \quad |f;g\rangle = -|g;f\rangle \qquad 5.12$$

**Theorem:** *The Pauli exclusion principle holds for fermions.*

The definition of the creation operator extends to $|x\rangle : \mathbb{H}^n \to \mathbb{H}^{n+1}$ by requiring that its action on each particle of an $n$ particle state is identical, and that it reduces to 5.1 in the restriction of $\mathbb{H}^n$ to the space of the $i$th particle. Thus $\forall x, y^i \in N, i = 1, ..., n$

$$|x\rangle : (|y^1\rangle, ..., |y^n\rangle) \to \frac{1}{\sqrt{n+1}}\left((|x\rangle, |y^1\rangle, ..., |y^n\rangle) + \kappa \sum_{i=1}^{n} (|y^i\rangle, |y^1\rangle, ..., |x\rangle, ..., |y^n\rangle)\right) \qquad 5.13$$

where $\kappa = 1$ for bosons, $\kappa = -1$ for fermions, and $|x\rangle$ appears in the $i+1$th position in the $i$th term of the sum. Normalisation is determined from 4.2 by observing that when all $x, y^i$ are distinct, the right hand side is the sum of $n+1$ orthonormal vectors. 5.13 is extended $\forall |f\rangle \in \mathbb{H}$ and $\forall |g\rangle \in \mathbb{H}^n$ by linearity.

**Definition:** *The space of physically realisable states $\mathscr{F} = \mathscr{F}(\mathbb{N}) \subset \mathscr{H}(\mathbb{N})$ is the subspace generated from $\mathbb{H}^0 = \{|\rangle\}$ by the action of creation operators (more strictly physical states are generated by interaction operators which will composed of creation and annihilation operators).*

**Definition:** *Notation for the elements of $\mathscr{F}$ is defined inductively.*

$$\forall |g\rangle \in \mathbb{H}^1, \forall |f\rangle \in \mathbb{H}^n \cap \mathscr{F} \quad |g;f\rangle = |g\rangle|f\rangle \in \mathbb{H}^{n+1} \cap \mathscr{F} \qquad 5.14$$

**Corollary:** $|g;f\rangle$ *is identified with the creation operator $\mathscr{H} \to \mathscr{H}$ given by $|g;f\rangle = |g\rangle|f\rangle$*

**Definition:** *The bra corresponding to $|g;f\rangle \in \mathbb{H}^{n+1}$ is $\langle g;f|$.*

**Theorem:** $\forall |x^i\rangle \in \mathbb{H}_0^1, i = 1, \ldots, n$

$$|x^1; x^2; \ldots; x^n\rangle = \frac{1}{\sqrt{n!}} \sum_\pi \varepsilon(\pi)(|x^{\pi(1)}\rangle, \ldots, |x^{\pi(n)}\rangle) \qquad 5.15$$

*where the sum runs over permutations $\pi$ of $(1,2,\ldots,n)$, $\varepsilon(\pi)$ is the sign of $\pi$ for fermions and $\varepsilon(\pi)=1$ for bosons.*

*Proof:* By induction, 5.15 holds for $n = 2$, by 5.6 and 5.7. Now suppose that 5.15 holds $\forall n < m \in \mathbb{N}$, Then, from definition 5.14

$$|x^1; x^2; \ldots; x^m\rangle = |x^1\rangle|x^2; \ldots; x^m\rangle$$

$$= \frac{1}{\sqrt{(m-1)!}} \sum_\pi \varepsilon(\pi)|x^1\rangle(|x^{\pi(2)}\rangle, \ldots, |x^{\pi(m)}\rangle)$$

by the inductive hypothesis. 5.15 follows from application of 5.13.

**Corollary:** $\forall |g\rangle, |f\rangle \in \mathbb{H}$ the creation operators obey the (anti)commutation relations

$$[|g\rangle, |f\rangle]_\pm = 0 \qquad 5.16$$

where, for fermions and bosons respectively

$$[x, y]_+ = \{x, y\} = xy + yx \text{ and } [x, y]_- = [x, y] = xy - yx$$

*Proof:* By definition 5.14, $\forall x^i \in \mathbb{N}, i = 1, \ldots, n$, $\forall |x\rangle, |y\rangle \in \mathbb{H}_0$

$$|x\rangle, |y\rangle|x^1; x^2; \ldots; x^n\rangle = |x; y; x^1; x^2; \ldots; x^n\rangle$$
$$= \kappa|y; x; x^1; x^2; \ldots; x^n\rangle \qquad \text{by 5.15}$$
$$= \kappa|x\rangle, |y\rangle|x^1; x^2; \ldots; x^n\rangle$$

But by definition the kets $|x^1; x^2; \ldots; x^n\rangle$ span $\mathscr{F}$. So by linearity

$$[|x\rangle, |y\rangle]_\pm = 0$$

5.16 follows from 5.10.

**Theorem:** $\forall |x^i\rangle, |y^i\rangle \in \mathbb{H}_0, i = 1, \ldots, n$

$$\langle y^1; \ldots; y^n | x^1; \ldots; x^n\rangle = \sum_\pi \varepsilon(\pi) \prod_{i=1}^n \langle y^i | x^{\pi(i)}\rangle \qquad 5.17$$

*Proof:* By 5.15 and 4.2

$$\langle y^1;\ldots;y^n \mid x^1;\ldots;x^n \rangle = \frac{1}{n!}\sum_{\pi'}\varepsilon(\pi')\sum_{\pi''}\varepsilon(\pi'')\prod_{i=1}^{n}\langle y^{\pi'(i)}\mid x^{\pi''(i)}\rangle$$

$$= \frac{1}{n!}\sum_{\pi'}\varepsilon(\pi')\sum_{\pi\pi'}\varepsilon(\pi\pi')\prod_{i=1}^{n}\langle y^{\pi'(i)}\mid x^{\pi\pi'(i)}\rangle$$

where we observe that $\forall$ permutations $\pi''$, $\pi'$, $\exists$ a permutation $\pi$ such that $\pi'' = \pi\pi'$. 5.17 follows since the sum over $\pi'$ contains $n!$ terms which are identical up to the ordering of the factors in the product.

**Corollary:** $\forall \mid g_i\rangle, \mid f_j\rangle \in \mathbb{H}, i,j = 1,\ldots,n$

$$\langle g_1;\ldots;g_n \mid f_1;\ldots;f_n \rangle = \sum_{\pi}\varepsilon(\pi)\prod_{i=1}^{n}\langle g_i\mid f_{\pi(i)}\rangle \qquad 5.18$$

*Proof:* By linearity, 5.9, and definition 5.14

**Theorem:** $\forall n \in \mathbb{N}$, such that $0 < n$, $(\mathscr{T}\cap \mathbb{H}^n) \subset (\mathscr{T}\cap \mathbb{H}^{n+1})$ *is an isomorphic embedding under the mapping* $\mathbb{H}^n \to \mathbb{H}^{n+1}$ *given by*

$$\forall x^i \in \mathbb{N}, i = 1,\ldots,n \quad \mid x^1;\ldots;x^n \rangle \to \mid\rangle\mid x^1;\ldots;x^n \rangle = \mid;x^1;\ldots;x^n \rangle \qquad 5.19$$

*Proof:* By 5.17 and 4.1

$$\langle ;y^1;\ldots;y^n \mid ;x^1;\ldots;x^n \rangle = \sum_{\pi\neq 1}\varepsilon(\pi)\langle\mid\rangle\prod_{i=2}^{n+1}\langle y^i\mid x^{\pi(i)}\rangle$$

$$= \langle y^1;\ldots;y^n \mid x^1;\ldots;x^n \rangle$$

since $\langle\mid\rangle = 1$ and using 5.17 again.

## 6 Annihilation Operators

In an interaction particles may be created, as described by creation operators, and particles may change state or be destroyed. The destruction of a particle in interaction is described by the action of an annihilation operator. A change of state of a particle can be described as the annihilation of one state and the creation of another, so a complete description of any process in interaction can be achieved through combinations of creation and annihilation operators. Annihilation operators incorporate the idea that it is impossible to tell which particle of given type has been destroyed in the interaction. They are defined by their action on a basis of $\mathscr{H}$, and their relationship to creation operators will be determined. The use of bras to denote annihilation operators is justified by the obvious homomorphism defined below in 6.2 with $n = 1$. Certain proofs of a routine nature have been omitted.

**Definition:** $\forall \mid x\rangle \in \mathbb{H}_0$ *the annihilation operator* $\langle x\mid:\mathbb{H}^n \to \mathbb{H}^{n-1}$ $\langle x\mid:\mid f\rangle \to \langle x\mid f\rangle \in \mathbb{H}^{n-1}$ *is given by* $\forall x^i \in \mathbb{N}, i = 1,\ldots,n$

$$\langle x\mid\mid\rangle = \langle x\mid\rangle \qquad 6.1$$

$$\langle x\mid(\mid x^1\rangle,\ldots,\mid x^n\rangle) = \frac{1}{\sqrt{n}}\sum_{i=1}^{n}\kappa^i\langle x\mid x^i\rangle(\mid x^1\rangle,\ldots,\mid x^{i-1}\rangle,\mid x^{i+1}\rangle,\ldots,\mid x^n\rangle) \qquad 6.2$$

The normalisation in 6.2 is determined by observing that when all $x, x^i$ are distinct, the right hand side is the sum of $n$ orthonormal vectors. $\kappa = 1$ for bosons and $\kappa = -1$ for fermions, and is determined by considering the result of the annihilation operator on a state of one particle in $\mathbb{H}^1 \subset \mathbb{H}^n \cap \mathscr{F}$, which is identical for all values of $n$ under the isomorphic embedding of 5.19. The annihilation operator for any ket is defined by linearity

$$\forall \, |f\rangle \in \mathbb{H} \quad \langle f|: \mathscr{F} \to \mathscr{F} \text{ is given by } \langle f| = \sum_{x \in \mathbb{N}} \langle f|x\rangle\langle x| \qquad 6.3$$

**Lemma:** $\forall |x\rangle, |x^1\rangle, |x^2\rangle \in \mathbb{H}_0^1$

$$\langle x|(|x^1\rangle, |x^2\rangle) = \frac{1}{\sqrt{2}}\langle x|x^1\rangle|x^2\rangle + \kappa\langle x|x^2\rangle|x^1\rangle \qquad 6.4$$

*Proof:* This is 6.2 with $n = 2$

**Theorem:** $\forall |y\rangle, |x^i\rangle \in \mathbb{H}_0, i = 1, \ldots, n$

$$\langle y|x^1; \ldots; x^n\rangle = \sum_{i=1}^{n} \kappa^i \langle y|x^i\rangle |x^1; \ldots; x^{i-1}; x^{i+1}; \ldots; x^n\rangle \qquad 6.5$$

*Proof:* By 5.15

$$\langle y \| x^1; \ldots; x^n\rangle = \langle y| \frac{1}{\sqrt{n!}} \sum_\pi \varepsilon(\pi)(|x^{\pi(1)}\rangle, \ldots, |x^{\pi(n)}\rangle)$$

$$= \frac{1}{\sqrt{n}} \frac{1}{\sqrt{n!}} \sum_{i=1}^{n} \kappa^i \langle y|x^i\rangle n \sum_{\pi \neq i} \varepsilon(\pi)(|x^{\pi(1)}\rangle, \ldots, |x^{\pi(n)}\rangle)$$

by 6.2, since for each value of $i \in \{1, \ldots, n\}$ there are $n$ permutations $\pi$ which are identical apart from the position of $i$. 6.5 follows by applying 5.15 again.

**Theorem:** $\forall |x^i\rangle, |y^i\rangle \in \mathbb{H}_0, i = 1, \ldots, n$

$$\langle y^n|\ldots\langle y^1 \| x^1; \ldots; x^n\rangle = \langle y^1; \ldots; y^n | x^1; \ldots; x^n\rangle \qquad 6.6$$

*Proof:* By induction, left to the reader.

**Corollary:** $\forall |x^i\rangle \in \mathbb{H}_0, i = 1, \ldots, n \quad \forall |f\rangle \in \mathbb{H}^n \cap \mathscr{F}$

$$\langle x^1; \ldots; x^n | f\rangle = \langle x^n| \ldots \langle x^1 \| f\rangle$$

*Proof:* Immediate from 6.6, by linearity. Hence, it is consistent to define:

**Definition:** $\forall |x^i\rangle \in \mathbb{H}_0, i = 1, \ldots, n$ the annihilation operator $\langle x^1; \ldots; x^n|: \mathscr{H} \to \mathscr{H}$ is given by

$$\langle x^1; \ldots; x^n| = \langle x^n| \ldots \langle x^1| \qquad 6.7$$

**Definition:** *On a complex vector space, $\mathscr{V}$, with a hermitian form, the hermitian conjugate $\phi^\dagger: \mathscr{V} \to \mathscr{V}$ of the linear operator $\phi: \mathscr{V} \to \mathscr{V}$ is defined such that $\forall f, g \in \mathscr{V}. (\phi^\dagger f, g) = (f, \phi g)$.*

**Definition:** $\forall |x^i\rangle \in \mathbb{H}_0, i = 1, \ldots, n$ *the creation operator $|x^1; \ldots; x^n\rangle: \mathscr{F} \to \mathscr{F}$ is the hermitian conjugate of the annihilation operator, $\langle x^1; \ldots; x^n|: \mathscr{F} \to \mathscr{F}$.*

$$\langle x^1; \ldots; x^n|^\dagger = |x^1; \ldots; x^n\rangle \qquad 6.8$$

*Proof:* From the definition, $\forall x^i, y^j \in \mathrm{N}, i = 1, \ldots, n, j = 1, \ldots, m \; \forall |f\rangle \in \mathscr{F}$,

$$\langle y^1;\ldots;y^n|\langle x^1;\ldots;x^n|^\dagger|f\rangle = \langle y^1;\ldots;y^n|\langle x^1;\ldots;x^n\|f\rangle$$
$$= \langle y^n|\ldots\langle y^1|\langle x^n|\ldots\langle x^1\|f\rangle$$
$$= \langle x^1;\ldots;x^n;y^1;\ldots;y^n\|f\rangle$$

by applying 6.7 three times. Thus $\langle x^1;\ldots;x^n|^\dagger$ is the map

$$\langle x^1;\ldots;x^n|^\dagger:|y^1;\ldots;y^n\rangle \rightarrow |x^1;\ldots;x^n;y^1;\ldots;y^n\rangle$$

which demonstrates 6.8.

**Corollary:** $\forall |g\rangle, |f\rangle \in \mathbb{H}$ *the annihilation operators obey the (anti)commutation relations.*

$$[\langle g|,\langle f|]_\pm = 0 \qquad 6.9$$

*Proof:* Straightforward from, 5.16, the (anti)commutation relations for creation operators.

**Theorem:** $\forall |g\rangle, |f\rangle \in \mathbb{H}$ *the creation operators and annihilation operators obey the (anti)commutation relations*

$$[\langle g|,|f\rangle]_\pm = \langle g|f\rangle \qquad 6.10$$

*Proof:* From 6.5 and using linearity, left to the reader

# 7 Interactions

In this treatment $\mathscr{F}$ is simply a labelling system and its construction has required no physics beyond the knowledge that we can measure the position of individual particles, and that we can measure the relative frequency of each result of a repeated measurement. The description of physical processes in terms of this labelling system requires a law describing the time evolution of states. Let $\mathrm{T} \subset \mathbb{N}$ be a finite discrete time interval such that any particle under study certainly remains in N for $x_0 \in \mathrm{T}$. Without loss of generality let $\mathrm{T} = [0, T)$. An interaction at time $t$ is described by an operator, $I(t):\mathscr{F} \rightarrow \mathscr{F}$. For definiteness we may take

$$\forall x^i \in \mathrm{N}, \forall n \in \mathbb{N}, \langle x^1;\ldots;x^n|I|x^1;\ldots;x^n\rangle = 0 \qquad 7.1$$

since otherwise there would be a component of *I* corresponding to the absence of interaction. At each time *t*, either no interaction takes place and the state $|f\rangle \in \mathscr{F}$ is unchanged, or an interaction, *I*, takes place. By the identification of the operations of vector space with weighted OR between uncertain possibilities, the possibility of an interaction at time *t* is described by the map $\mathscr{F} \rightarrow \mathscr{F}$

$$|f\rangle \rightarrow \mu(1 - iI(t))|f\rangle$$

where $\mu$ is a scalar chosen to preserve the norm, as required by the probability interpretation. Thus the law of evolution of the ket from time *t* to time $t + 1$ is

$$|f\rangle_{t+1} = \mu(1 - iI(t))|f\rangle_t \qquad 7.2$$

It is straightforward to see that 7.2 is in some sense approximated by the time evolution equation $|f\rangle_{t+1} = \mu e^{-iI(t)}|f\rangle_t$ found in, for example, lattice field theory [13]. Lattice field theory does not describe the model of simple particle interactions considered here, so there is motivation for a somewhat modified treatment. Observe that Hilbert space is information space at some time *t*, $\mathscr{F} = \mathscr{F}(t)$, so the interaction is

a map

$$I(t); \mathscr{F}(t) \to \mathscr{F}(t+1)$$

Since $I(t)$ is a map from one Hilbert space to another we cannot talk of it being self adjoint, or of the spectrum of $1 - iI(t)$. Although we can identify the Hilbert spaces at different times using the natural homomorphism defined by the basis of position kets, when we do so we cannot use a linear operator $I(t)$ on the resultant space to physically represent an interaction. Linearity is normally imposed since when $I(t)$ acts on a state prior history should not be relevant. But a more careful analysis suggests that $I(t)$ is not strictly linear, because linearity would dictate that an the action of $I(t)$ on a particle created by $I(t)$ at time $t$ should be the same as its action on a particle previously created and evolving to the same ket, so that a particle can physically interact twice in the same instant. This appears physically meaningless and we regularise $I(t)$ by imposing the condition that a particle cannot be annihilated at the instant of its creation. In other words $I(t)$ cannot act on the results of itself. This regularisation only makes sense if the unit, $\chi$, describing one instant of time is actually a fundamental physical parameter of the universe equal to lattice spacing. Thus we have

$$I^2(t) = 0 \qquad 7.3$$

In other respects $I(t)$ is linear. The removal of products describing the annihilation of a particle at the instant of its creation, as in 7.5 is most naturally done by normal ordering. In practice this is largely academic, since when we find the perturbation expansion as an iterative solution of 7.2 terms containing $I^2(t)$ are already excluded. We will see that the exclusion of these terms leads directly to a finite "renormalised" perturbation expansion. It might be thought that 7.3 would prevent an interaction taking place in each instant, and so would prevent interaction altogether but this is only the case if there is also an observation in each instant. We may regard this as a limiting instance of a quantum Zeno effect, which is known in quantum mechanics to stop interaction under conditions of continuous observation [10]

By 7.2 preservation of the norm implies that $\forall |f\rangle \in \mathscr{F}$

$$\langle f | f \rangle = \langle f | (1 + iI^\dagger)\bar{\mu}\mu(1-iI) | f \rangle \qquad 7.4$$

$$= |\mu|^2 (\langle f | f \rangle + \langle f | I^\dagger I | f \rangle + i \langle f | I^\dagger - I | f \rangle) \qquad 7.5$$

Normal ordering implies that $\langle f | (I^\dagger I) | f \rangle = 0$ So

$$\frac{\langle f | I^\dagger - I | f \rangle}{\langle f | f \rangle} = i \frac{1 - |\mu|^2}{|\mu|^2} \qquad 7.6$$

7.6 has a straightforward solution with $|\mu|^2 = 1$. and $I = I^\dagger$. Although strictly non-linearity implies that $I$ is not hermitian, $I^2(t)$ does not appear in the physical model, as discussed above, and we may treat $I$ as hermitian and will refer to it as such. 7.2 can be interpreted literally as meaning that in each instant particle either interacts or does not interact. In the latter case the state remains the same and is multiplied by a phase, $\mu = e^{-iE}$, $E \in \mathbb{R}$ so that 7.2 reduces to

$$|f\rangle_{t+1} = e^{-iE}|f\rangle_t \qquad 7.7$$

7.7 is a geometric progression with solution

$$|f\rangle_t = e^{-iEt}|f\rangle_0 \qquad 7.8$$

## 8 Covariance

Clearly the use of a discrete lattice requires that we redefine Lorentz transformation. The lattice is invariant because transformation introduces a new lattice, aligned with the new coordinate axes and retaining the parameters $\chi$ and $\nu$. $\nu$ is an arbitrary observer choice. If lattice size, $\nu$, is changed then momentum space is redefined; this must be done in co-ordinate space. The same is true of any change of co-ordinate system, such as to curvilinear co-ordinates. As seen in section 7, for a useful time evolution equation $\chi$ is a fundamental physical constant, and is the same in all reference frames. We will continue to use $\chi = 1$. We only require local invariance as in general relativity, and we seek to carry out transformations only when all the matter under study is contained in the reference frames both before and after transformation. We require a proscription relating predictions made in one lattice with predictions made in another. From 3.6 the general solution of 7.7 at time $t$ is

$$\langle x, t | f \rangle = \left(\frac{1}{2\pi}\right)^{\frac{3}{2}} \int_M d^3p \, \langle p | f \rangle \, e^{-i(Et - x \cdot p)} \qquad 8.1$$

Phase, $\mu = e^{iE}$, is arbitrary, but we seek a covariant solution and we fix $\mu$ by defining:

**Definition:** $E = p_0$ *is the time like component of a vector* $p = (E, \boldsymbol{p})$. *E is called energy.*

Only the integrand in 8.1 is covariant. The integral operator is defined in an identical manner in all reference frames, and is invariant. It will be found that $E$ is conserved in measured states and can be identified with classical energy. By definition we have a mass shell condition $m^2 = E^2 - \boldsymbol{p}^2$ where $m$ is a constant, known as bare mass. It follows immediately that elementary particles obey the Klein-Gordon equation. In this treatment the Klein-Gordon equation is an identity based on the definition of energy as the time component of an equation of motion, and will not be treated as an equation of motion.

Although $\langle x, t | f \rangle$ is discrete in $x$ and $t$, on a macroscopic scale it appears continuous. 8.1 can be embedded into a continuous function $f: \mathbb{R}^4 \to \mathbb{C}$, called the wave function defined by

$$\forall x \in \mathbb{R}^4 \quad f(x) = \left(\frac{1}{2\pi}\right)^{\frac{3}{2}} \int_M d^3p \, \langle p | f \rangle \, e^{-ix \cdot p} \qquad 8.2$$

So we have

$$\forall \boldsymbol{x} \in \mathrm{N}, \forall t \in \mathrm{T} \quad \langle x | f \rangle = \langle t, \boldsymbol{x} | f \rangle = f(t, \boldsymbol{x}) = f(x) \qquad 8.3$$

8.2 is not manifestly covariant, but under reasonable conditions it is invariant for physically realisable states and transformations. To see this we observe that since energy is conserved in physical measurement momentum is always bounded by the total energy of a system (it may not be strictly possible to bound the support in both momentum space and coordinate space, but it is possible to do so to the accuracy of experiment). The probability of finding a momentum above this bound is zero, and we assume that for physically realisable states the discrete representation of $\langle p | f \rangle$ has support which is bounded below $\pi/\chi$ in each component of momentum. The bound depends on the system under consideration, but without wishing to specify it, we observe that, provided that the discrete unit of time is sufficiently small, it is always much less than $\pi/\chi$. Physically meaningful Lorentz transformation cannot boost it beyond this value because physical reference frames are defined with respect to macroscopic matter. A realistic Lorentz transformation means that macroscopic matter has been physically boosted by the amount of the transformation. If lattice spacing is $Gm_e/c^3$ a boost in the order of $\pi$ would require an energy of $9 \times 10^{14}$ solar masses per kilogram of matter to be boosted, which may be thought unrealistic. Thus we can ensure covariance by imposing the condition that all physical momentum space wave functions have a representation in a subset of M bounded by some realistic energy level (this bound will not affect the calculation of loops in Feynman diagrams, since these are not observable states). Then we remove the non-physical periodic property of $\langle p|f\rangle$ by replacing

$$\Theta_M(p)\langle p|f\rangle \to \langle p|f\rangle$$

where $\Theta_M(p) = 1$ if $p \in M$ and $\Theta_M(p) = 0$ otherwise. 8.2 is then replaced with

$$f(x) = \left(\frac{1}{2\pi}\right)^{\frac{3}{2}} \int_{\mathbb{R}^3} d^3p \, \langle p|f\rangle \, e^{-ix \cdot p} \qquad 8.4$$

For any ket, there is a unique momentum space function $\langle p|f\rangle$ defined by 3.4, and a unique wave function defined by 8.4. So in each coordinate system $N \in \mathcal{N}$ there is a homomorphism between $\mathbb{H}$ and the vector space of wave functions with the hermitian product defined by 2.5

$$\langle g|f\rangle = \sum_{x \in N} \overline{g(x)} f(x) \qquad 8.5$$

Clearly Lorentz transformation cannot be applied directly to a discrete co-ordinate system, but it can be applied to the wave function, 8.4. Then 8.3 defines the position function, and hence a ket, by the restriction of the wave function to the transformed co-ordinate system, $N' \in \mathcal{N}$, at integer time. Let $|x\rangle$ be a state of a particle at definite position $x$ in the lattice at some time $x^0$. Then, from 3.11, the Lorentz transformation is

$$|\Lambda x\rangle = \int_{\mathbb{R}^3} d^3p |p\rangle\langle p|\Lambda x\rangle = \int_{\mathbb{R}^3} d^3p |p\rangle\langle \Lambda' p|x\rangle$$

So we have for $\Lambda$

$$\int_{\mathbb{R}^3} d^3p |\Lambda p\rangle\langle p| = \Lambda = \int_{\mathbb{R}^3} d^3p |p\rangle\langle \Lambda' p| \qquad 8.6$$

We impose a new co-ordinate system at time $t'$ after transformation by restricting the wave function to points $x'$ in a new cubic lattice N'. Then the transformed state is the restriction to the new lattice, i.e

$$|\Lambda x\rangle = \sum_{N'} |x'\rangle\langle x'|\Lambda x\rangle$$

8.6 gives

$$|\Lambda x\rangle = \int_{\mathbb{R}^3} d^3p \sum_{N'} |x'\rangle\langle x'|p\rangle\langle \Lambda' p|x\rangle$$

$|\Lambda x\rangle$ is not an eigenstate of position in N'; if a measurement of position were done in N' and we were then to transform back to N the state would no longer be $|x\rangle$. Thus the operators for position in different frames N and N' do not commute. But if no measurement is done, we can transform straight back and recover $|x\rangle$, showing that there is no problem with lack of unitarity under Lorentz transformation.

*Proof:*

$$\Lambda'|\Lambda y\rangle = \sum_{x \in N'} \int_{\mathbb{R}^3} d^3p \Lambda'|x'\rangle\langle x'|p\rangle\langle \Lambda' p|y\rangle$$

Restrict to the original coordinates

$$\begin{aligned}\Lambda'|\Lambda\,y\,\rangle &= \sum_{x\in\mathrm{N}}\sum_{x'\in\mathrm{N}'}\int_{\mathbb{R}^3}d^3p\,|x\,\rangle\langle x|\Lambda'x'\rangle\langle x'|p\,\rangle\langle \Lambda'\,p\,|\,y\,\rangle \\ &= \sum_{x\in\mathrm{N}}\sum_{x'\in\mathrm{N}'}\int_{\mathbb{R}^3}d^3p\,|x\,\rangle\langle \Lambda\,x|x'\rangle\langle x'|\Lambda p\,\rangle\langle p\,|\,y\,\rangle \\ &= \sum_{x\in\mathrm{N}}\int_{\mathbb{R}^3}d^3p\,|x\,\rangle\langle \Lambda\,x|\Lambda p\,\rangle\langle p\,|\,y\,\rangle \\ &= \sum_{x\in\mathrm{N}}\int_{\mathbb{R}^3}d^3p\,|x\,\rangle\langle x|p\,\rangle\langle p\,|\,y\,\rangle \\ &= |y\,\rangle \end{aligned}$$

## 9  Wave Mechanics

Wave functions are not restricted to $\mathcal{L}^2$, but, in any reasonable definition of an integral, 8.5 is approximated by the hermitian product in $\mathcal{L}^2$ whenever $f$ and $g$ are in $\mathcal{L}^2$ and the spacing of the lattice is small. The law for the time evolution of the wave function is given by differentiating 8.2

$$i\partial_0 f = Ef \qquad 9.1$$

7.7 is obtained by integrating 9.1 over one time interval. Thus, in the restriction to integer values, 9.1 is identical to 7.7, the difference equation for a non-interacting particle. It is therefore an expression of the same relationship or law. As an equation of the wave function, the right hand side of 9.1 is a scalar, whereas the left hand side is the time component of a vector whose space component is zero. So 9.1 is not manifestly covariant. A covariant equation which reduces to 9.1 requires the scalar product between the vector derivative, $\partial$, and the wave function and has the form, for some vector operator $\Gamma$

$$i\partial \cdot \Gamma f = mf \qquad 9.2$$

Then the time evolution of the position function in any reference frame N is the restriction of the solution of 9.2 to N at time $t \in \mathrm{T}$. As discovered by Dirac [4], there is no invariant equation in the form of 9.2 for scalar $f$ and the theory breaks down. To rectify the problem a spin index is added to N

$$\mathrm{N}_S = \mathrm{N} \otimes S \qquad \text{for } v \in \mathbb{N}$$

where $S$ is a finite set of indices. When there is no ambiguity we write $\mathrm{N} = \mathrm{N}_S$, and the constructions of the vector spaces, $\mathbb{H}$, $\mathcal{H}$ and $\mathcal{T}$ go through as before.

When we wish to make the spin index explicit we write $|x\rangle = |x, \alpha\rangle = |x\rangle_\alpha$ normalised by 2.1

$$\forall (x, \alpha), (y, \beta) \in \mathrm{N}_S \quad \langle x, \alpha|y, \beta\rangle = \langle x|y\rangle_{\alpha\beta} = \delta_{xy}\delta_{\alpha\beta} \qquad 9.3$$

The wave function acquires a spin index

$$f(x) = f_\alpha(x) = \langle x|f\rangle_\alpha \qquad 9.4$$

and the braket becomes

$$\langle g|f\rangle = \sum_{x\in\mathrm{N}_S}\langle g|x\rangle\langle x|f\rangle = \sum_{(\mathbf{x},\alpha)\in\mathrm{N}_S}\overline{g_\alpha(x)}f_\alpha(x) \qquad 9.5$$

It is now possible to find a covariant equation which reduces to 9.1 in the particle's reference frame, namely the Dirac equation,

$$i\partial \cdot \gamma f(x) = mf(x) \qquad 9.6$$

Another possibility is that $f$ is a vector and that 9.1 is a representation of a vector equation (with $\Gamma = 1$)

$$i\partial \cdot f(x) = 0 \qquad 9.7$$

9.7 is understood as the equation of motion of a vector particle which is only ever created or destroyed in interaction, and for which there is no interval of proper time between creation and annihilation (zero proper time for the emission and absorption of a light beam is a familiar result from special relativity). The norm is intended to generate physically realisable predictions of probability, and must be both invariant and positive definite. It is given by

$$\langle f | f \rangle = \sum_{(\mathbf{x}, \alpha) \in N_S} \overline{f_\alpha(x)} f_\alpha(x) \qquad 9.8$$

If $f$ transforms as a space-time vector, 9.8 is only invariant if 9.3 is replaced by the definition

$$\forall (\mathbf{x}, \alpha), (\mathbf{y}, \beta) \in N \quad \langle \mathbf{x}, \alpha | \mathbf{y}, \beta \rangle = \langle \mathbf{x} | \mathbf{y} \rangle_{\alpha\beta} = \eta(\alpha)\delta_{xy}\delta_{\alpha\beta} \qquad 9.9$$

where $\eta$ is given by

$$\eta(0) = -1 \text{ and } \eta(1) = \eta(2) = \eta(3) = 1.$$

We will use the summation convention for repeated spin indices, but not the convention of raising and lowering indices. The factor -1 is implicit in summing the zeroeth index for vectors, so 9.5 and 9.8 are retained. 9.9 is invariant, but not positive definite, as required by a norm. The definition of the braket in terms of probability implies that any vector particles have a positive definite norm for physical states, so only space-like polarisation of massless vector particles can be observed physically. Other states are permitted, and are required if interactions are to give correct physical predictions, but we can only discuss the probability of observing them if there is positive definite norm.

## 10  Dirac Particles

The solution to 9.6 is

$$f_\alpha(x) = \left(\frac{1}{2\pi}\right)^{\frac{3}{2}} \sum_{r=1}^{2} \int_{\mathbb{R}^3} d^3p \; F(p, r) u_\alpha(p, r) \; e^{-ix \cdot p} \qquad 10.1$$

where $p$ satisfies the mass shell condition and $u$ is a Dirac spinor, having the form

$$u(p, r) = \sqrt{\frac{p_0 + m}{2p_0}} \begin{bmatrix} \zeta(r) \\ \frac{\sigma \cdot p}{p_0 + m} \zeta(r) \end{bmatrix} \quad \text{for } r = 1,2 \qquad 10.2$$

where $\zeta$ is a two-spinor normalised so that

$$\overline{\zeta}_\alpha(r)\zeta_\alpha(s) = \delta_{rs} \qquad 10.3$$

and $\sigma = (\sigma_1, \sigma_2, \sigma_3)$ are the Pauli spin matrices.

It is routine to show the spinor normalisation

$$\bar{u}_\alpha(\boldsymbol{p}, r) u_\alpha(\boldsymbol{p}, s) = \delta_{rs}$$

$F(\boldsymbol{p},r)$ is the momentum space wave function given by inverting 10.1 at $x_0 = 0$

$$F(\boldsymbol{p}, r) = \left(\frac{\chi}{2\pi}\right)^{\frac{3}{2}} \sum_{(\boldsymbol{x}, \alpha) \in N} f_\alpha(0, \boldsymbol{x}) \bar{u}_\alpha(\boldsymbol{p}, r) e^{i\boldsymbol{x} \cdot \boldsymbol{p}} \quad 10.4$$

**Definition:** *$p_0$ is the energy of a state with momentum $\boldsymbol{p}$. $p = (p_0, \boldsymbol{p})$ is called energy-momentum; $p_0$ will later be identified with classical energy.*

**Definition:** *With the Dirac $\gamma$-matrices as defined in the literature the Dirac adjoint is*

$$\hat{u} = \bar{u}\gamma^0$$

**Lemma:** *The $\gamma$-matrices obey the relations*

$$\gamma^0\gamma^0 = 1 \text{ and } \gamma^0\gamma^{\alpha\dagger}\gamma^0 = \gamma^\alpha \quad 10.5$$

*Proof:* These are familiar matrix equations and the proof is left to the reader

**Lemma:** *In this normalisation Dirac spinors obey the following relations*

$$(p \cdot \gamma - m)u(\boldsymbol{p}, r) = 0 = \hat{u}(\boldsymbol{p}, r)(p \cdot \gamma - m) \quad 10.6$$

$$\hat{u}_\alpha(\boldsymbol{p}, r)u_\alpha(\boldsymbol{p}, s) = \delta_{rs}\frac{m}{p_0}$$

$$\sum_{r=1}^{2} u_\alpha(\boldsymbol{p}, r)\hat{u}_\beta(\boldsymbol{p}, r) = \left(\frac{p \cdot \gamma + m}{2p_0}\right)_{\alpha\beta} \quad 10.7$$

*Proof:* These are familiar spinor relations renormalised and the proof is left to the reader.

This normalisation is consistent the definition of ket space in the reference frame of an individual observer and leads to some simplification of the formulae. Wave functions are non-physical and it is not necessary to use the invariant integral.

The most fundamental representation of the discrete equation of motion, 7.2, is most readily understood as describing the interactions of a particle in the proper time of that particle. This being so there is no way to say that a particles proper time cannot become reversed with respect to macroscopic matter. The treatment of the antiparticle modifies the Stückelberg-Feynman [19],[6] interpretation by considering the mass shell condition. A sign is lost in the mass shell condition, due to the squared terms, but a time-like vector with a negative time-like component has a natural definition of $m < 0$. So permissible solutions of the Dirac equation, 9.6, have positive energy $E = p_0 > 0$ when $m$ is positive and negative energy when $m$ is negative. Complex conjugation reverses time while maintaining the probability relationship, 2.2, and restores positive energy, and we also change the sign of mass, $m \to -m$. Thus, given that no interaction takes place, the ket for a Dirac particle in its own reference frame evolves according to 7.8, for both $m > 0$ and $m < 0$. But the negative energy solution is transformed and satisfies

$$i\partial \cdot \bar{\gamma} f(x) = -m f(x) \quad 10.8$$

where $\bar{\gamma}$ is the complex conjugate, $\bar{\gamma}^j_{\alpha\beta} = \overline{\gamma^j_{\alpha\beta}}$. Although this is a slightly different from the positron wave function cited in e.g.[1] the treatments will be reconciled in the definition of the field operators. The solution to 10.8 is the wave function for the antiparticle

$$f(x) = \left(\frac{1}{2\pi}\right)^{\frac{3}{2}} \sum_{r=1}^{2} \int_{\mathbb{R}^3} d^3p \, F(p,r) \bar{v}(p,r) \, e^{-ix \cdot p} \qquad 10.9$$

where $p$ satisfies the mass shell condition, and $\bar{v}$ is the complex conjugate of the Dirac spinor.

$$v(p, r) = \sqrt{\frac{p_0 + m}{2p_0}} \begin{bmatrix} \dfrac{\sigma \cdot p}{p_0 + m} \zeta(r) \\ \zeta(r) \end{bmatrix} \qquad \text{for } r = 1,2$$

10.9 is the complex conjugate of the negative energy solution of the Dirac equation. The spinor has the normalisation

$$\bar{v}_\alpha(p, r) v_\alpha(p, s) = \delta_{rs}$$

$F(p,r)$ is the momentum space wave function given by

$$F(p, r) = \left(\frac{1}{2\pi}\right)^{\frac{3}{2}} \sum_{(x, \alpha) \in N} f_\alpha(0, x) v_\alpha(p, r) e^{ix \cdot p} \qquad 10.10$$

**Lemma:** *In this normalisation the Dirac spinors obey the following relations*

$$(p \cdot \gamma + m) v(p, r) = 0 = \hat{v}(p, r)(p \cdot \gamma + m)$$

$$\hat{v}_\alpha(p, r) v_\alpha(p, s) = \delta_{rs} \frac{m}{p_0} \qquad 10.11$$

$$\sum_{r=1}^{2} v_\alpha(p, r) \hat{v}_\beta(p, r) = \left(\frac{p \cdot \gamma - m}{2p_0}\right)_{\alpha\beta} \qquad 10.12$$

*Proof:* These are familiar spinor relations renormalised and the proof is left to the reader.

## 11  The Photon

We require a solution to the equation of motion, 9.7. By definition of energy (section 8) every particle obeys a Klein-Gordon equation. Positive definite norm implies that the only possibility has zero mass (massive vector bosons are allowed but no probability amplitude exists for them and only their decay products may be directly observed). The wave function for the photon is

$$f_\alpha(x) = \left(\frac{\chi}{2\pi}\right)^{\frac{3}{2}} \int_{\mathbb{R}^3} d^3p \, F(p, r) w_\alpha(p, r) \, e^{-ix \cdot p} \qquad 11.1$$

where
i.   $p^2 = 0$                         (from the Klein-Gordon equation with zero mass)
ii.  $w$ are orthonormal vectors given by
    a) time-like component:                 $w(p, r) = (1, \mathbf{0})$
    b) space-like components: for $r = 1,2,3$ $w(p, r) = (0, \mathbf{w}(p, r))$ are such that $\mathbf{w}(p, 3) = \mathbf{p}/p_0$ is longitudinal and $\mathbf{w}(p, r) \cdot \mathbf{w}(p, s) = \delta_{rs}$ so $\mathbf{w}(p,1)$ and $\mathbf{w}(p,2)$ are transverse

iii. *F* is such that the photon cannot be polarised in the longitudinal and time-like spin states, i.e.

$$F(p,0) = F(p,3) \qquad 11.2$$

*Proof:* With the above definitions

$$p.w(p,3) = p_0 = -p.w(p,0) \text{ and } p.w(p,1) = p.w(p,2) = 0$$

So that differentiating 11.1

$$i\partial \cdot f(x) = \left(\frac{1}{2\pi}\right)^{\frac{3}{2}} \sum_{r=0}^{3} \int_{\mathbb{R}^3} d^3p \ F(p,r)p \cdot w_\alpha(p,r) \ e^{-ix \cdot p}$$

$$= \left(\frac{1}{2\pi}\right)^{\frac{3}{2}} \sum_{r=0}^{3} \int_{\mathbb{R}^3} d^3p \ p_0(F(p,3) - F(p,0))e^{-ix \cdot p}$$

$$= 0$$

by 11.2. This establishes that 11.1 is the solution to 9.7

$F(p,r)$ is the momentum space wave function given by inverting 11.1 at $x_0 = 0$

$$F(p,r) = \left(\frac{\chi}{2\pi}\right)^{\frac{3}{2}} \eta(r) \sum_{(x,\alpha) \in N} f_\alpha(0,x) w_\alpha(p,r) e^{ix \cdot p} \qquad 11.3$$

## 12 Plane Wave States

**Definition:** $\forall x_0 \in T$ *plane wave states* $|p, r\rangle = \mathbb{H}$ *are defined by the wave functions*

$$\langle x | p, r \rangle = \left(\frac{1}{2\pi}\right)^{\frac{3}{2}} u(p,r) e^{-ix \cdot p} \text{ for the Dirac particle} \qquad 12.1$$

$$\langle x | p, r \rangle = \left(\frac{1}{2\pi}\right)^{\frac{3}{2}} \bar{v}(p,r) e^{-ix \cdot p} \text{ for the antiparticle, and} \qquad 12.2$$

$$\langle x | p, r \rangle = \left(\frac{1}{2\pi}\right)^{\frac{3}{2}} w(p,r) e^{-ix \cdot p} \text{ for the photon.} \qquad 12.3$$

**Theorem:** *(Newton's first law) In an inertial reference frame, an elementary particle in isolation has a constant momentum space wave function.* $\forall | f \rangle \in \mathbb{H}$

$$|f\rangle = \sum_r \eta(r) \int_{\mathbb{R}^3} d^3p |p, r\rangle \langle p, r | f \rangle \qquad 12.4$$

*Proof:* Clearly plane waves are solutions of 9.6, 10.8 and 9.7 so they describe the evolution of states in isolation. For each of the Dirac particle, antiparticle and photon, by 2.6, $\forall | f \rangle \in \mathbb{H}$, $\forall x_0 \in T$

$$\langle p, r | f \rangle = \sum_{x \in N} \langle p, r | x \rangle' \langle x | f \rangle \qquad 12.5$$

Substituting 12.1, 12.2 and 12.3 in 12.5, with $x_0 = 0$, and examining 10.4, 10.9 and 11.3 reveals

$$\langle p, r | f \rangle = F(p, r) \qquad \text{for the Dirac particles, and} \qquad 12.6$$

$$\langle p, r | f \rangle = \eta(r) F(p, r) \quad \text{for the photon.} \qquad 12.7$$

**Corollary:** *The time evolution of the position function of a particle in isolation is,* $\forall |f\rangle \in \mathbb{H}$

$$\langle x|f\rangle = \sum_r \eta(r) \int_{\mathbb{R}^3} d^3p \langle x|p,r\rangle\langle p,r|f\rangle \qquad 12.8$$

where $r = 0\text{-}3$ for photons, and $r = 1\text{-}2$ for Dirac particles ($\eta$ is redundant for a Dirac particle).
*Proof:* Substituting 12.6 and 12.1 into 10.1, 12.6 and 12.2 into 10.9, and 12.7 and 12.3 into 11.1 gives, in each case, 12.8.

**Corollary:** *The resolution of unity*

$$\sum_r \eta(r) \int_{\mathbb{R}^3} d^3p |p,r\rangle\langle p,r| = 1 \qquad 12.9$$

*Proof:* 12.4 is true for all $|f\rangle \in \mathbb{H}$.

**Corollary:** *The braket has the time invariant form*

$$\langle g|f\rangle = \sum_r \eta(r) \int_{\mathbb{R}^3} d^3p \langle g|p,r\rangle\langle p,r|f\rangle \qquad 12.10$$

*Proof:* Immediate from 12.9

**Theorem:** $\langle q,s|p,r\rangle$ *is a delta function on the test space of momentum space wave functions*

$$\langle q,s|p,r\rangle = \eta(r)\delta_{rs}\delta(p-q) \qquad 12.11$$

*Proof:* From 12.10, for plane wave $|q,s\rangle$

$$\langle q,s|f\rangle = \sum_r \eta(r) \int_{\mathbb{R}^3} d^3p \langle q,s|p,r\rangle\langle p,r|f\rangle$$

**Corollary:** *The braket for the photon is positive definite, as required by the probability interpretation.*
*Proof:* By 11.2 and 12.7 the time-like ($r = 0$) and longitudinal ($r = 3$) states cancel out in 12.10 and for photons as well as Dirac particles 12.10 reduces to

$$\langle g|f\rangle = \sum_{r=1}^{2} \int_{\mathbb{R}^3} d^3p \langle g|p,r\rangle\langle p,r|f\rangle \qquad 12.12$$

**Theorem:** *(Gauge invariance). Let $g$ be an arbitrary solution of $\partial^2 g = 0$. Then observable results are invariant under gauge transformation of the photon wave function given by*

$$f_\alpha(x) \to f_\alpha(x) + \partial_\alpha g(x) \qquad 12.13$$

*Proof:* It follows from 12.12 that the braket is invariant under the addition of a (non-physical) light-like polarisation state, known as a gauge term. Let $G(p)$ be an arbitrary function of momentum. The general solution for $g$ is

$$g = \int_{\mathbb{R}^3} d^3p\, e^{-ip \cdot x} G(p)$$

where $p^2 = 0$. Then

$$\partial_\alpha g = \int_{\mathbb{R}^3} d^3p\, p_0(w(p,0) + w(p,3)) e^{-ip \cdot x} G(p)$$

is equivalent to a light like polarisation states, and has no effect on the braket. $\partial_\alpha g$ is known as a gauge term,

and has no physical meaning. It follows from 12.12 that light-like polarisation cannot be determined from experimental results. Although their value is hidden by the gauge term, the time-like and longitudinal polarisation states cannot be excluded, and we will see that they contribute to the electromagnetic force.

**Theorem:** *Space-time translation by displacement, z, of the co-ordinate system such that the particle remains in* N, *is equivalent to multiplication of the momentum space wave function by* $e^{ip \cdot z}$.

*Proof:* Using 12.6 and/or 12.7 in 12.4.

$$\langle x | f \rangle = \sum_r \int_{\mathbb{R}^3} d^3 p \, F(p, r) \langle x | p, r \rangle \qquad 12.14$$

Under a space-time translation, $z$, by 12.1, 12.2 and 12.3 we have,

$$\langle x - z | f \rangle = \sum_r \int_{\mathbb{R}^3} d^3 p \, F(p, r) e^{ip \cdot z} \langle x | p, r \rangle \qquad 12.15$$

as required.

## 13 Field Operators

**Definition:** *A partial field* $\Psi = \Psi(N)$ *is a family of mappings* $\Psi(N): \mathbb{R}^4 \otimes S \to \mathcal{T}(N)$, *where S is the set of spin indices introduced in section 8, the elements of* $\mathcal{T}(N)$ *are regarded as operators.*

**Definition:** *The partial field of creation operators for a particle in interaction is* $|\underline{x, \alpha}\rangle$

$$\forall (x, \alpha) = (x_0, \boldsymbol{x}, \alpha) \in \mathbb{R}^4 \otimes S. \quad |\underline{x, \alpha}\rangle: \mathcal{T} \to \mathcal{T} \qquad 13.1$$

We will find that photons are not created in eigenstates of position so we do not in general have $\forall x \in N, |\underline{x, \alpha}\rangle = |x, \alpha\rangle$. $|\underline{x, \alpha}\rangle$ will be found for each particle.

**Definition:** *Let* $|\underline{\alpha}\rangle = |\underline{0, \alpha}\rangle$ *be the operator for the creation a particle at the origin.*

**Definition:** *By 6.8, the annihilation operator* $\langle \underline{x, \alpha}|: \mathcal{T} \to \mathcal{T}$ *is the hermitian conjugate.*

**Theorem:** *The creation operator* $|\underline{x, \alpha}\rangle: \mathcal{T} \to \mathcal{T}$ *for a particle at* $(x, \alpha) = \mathbb{R}^4 \otimes S$ *is given by*

$$|\underline{x, \alpha}\rangle = \sum_r \eta(r) \int_M d^3 p \langle p, r | \underline{\alpha} \rangle e^{ip \cdot x} | p, r \rangle \qquad 13.2$$

*Proof:* By the resolution of unity, 12.9, $|\underline{x, \alpha}\rangle: \mathcal{T} \to \mathcal{T}$ is given by

$$|\underline{x, \alpha}\rangle = \sum_r \eta(r) \int_M d^3 p \langle p, r | \underline{x, \alpha} \rangle | p, r \rangle \qquad 13.3$$

The momentum space bound has to be restored because we do not have a momentum space wave function with bounded support, but an operator on Hilbert space. By the principle of homogeneity space-time translation maps the creation operators appearing in interactions into each other. Then, by 12.15,

$$\langle p, r | \underline{x, \alpha} \rangle = \langle p, r | \underline{\alpha} \rangle e^{ip \cdot x} \qquad 13.4$$

13.2 follows by substituting 13.4 into 13.3.

**Definition:** *The derivative of the creation and annihilation operators is defined by differentiating 13.2.*

$$\partial |\underline{x}, \underline{\alpha}\rangle = |\partial \underline{x}, \underline{\alpha}\rangle = \sum_{r=0}^{3} \eta(r) \int_M d^3 p \langle p, r | \underline{\alpha}\rangle i p e^{ip \cdot x} | p, r\rangle \qquad 13.5$$

There may be a number of different types of interaction, described by $I_j: \mathscr{F} \to \mathscr{F}$, where $j$ runs over an index set. Let $e_j \in \mathbb{R}$ be the coupling constant for the interaction $I_j$. Only one type of interaction takes place at a time, but there is uncertainty about which. Under the identification of addition with quantum logical OR, the interaction operator $I(x_0): \mathscr{F} \to \mathscr{F}$ introduced in section section 7, is

$$I = \sum_j e_j I_j$$

$I$ is hermitian, and each $I_j$ is independent by definition, so each $I_j$ is hermitian 9 (up to regularisation).

**Definition:** *In any finite discrete time interval,* T, *for each type of interaction, an operator,*

$$H(x): \mathscr{F} \to \mathscr{F},$$

*describes the interaction taking place at* $x = (x_0, \boldsymbol{x}) \in T \otimes N$, $H(x)$ *is called interaction density.*

The principle of homogeneity implies that $H(x)$ is the same, up to homomorphism, and has equal effect on a matter anywhere in N and for all times in T. $I_j$ describes equal certainty that a particle interacts anywhere in N, so by the identification of addition with quantum logical OR, $I_j$ can be written as a sum

$$I_j(x_0) = \sum_{\boldsymbol{x} \in N} H(x_0, \boldsymbol{x}) = \sum_{\boldsymbol{x} \in N} H(x) \qquad 13.6$$

The sum in 13.6 is over space, but not necessarily over the spin index. Without loss of generality $H(x)$ is hermitian. By the definition of multiparticle space as a direct product (section 4), $H(x)$ can be factorised as a product of Hermitian operators, $J_\gamma(x)$, where $\gamma$ runs over the particles in the interaction

$$H(x) = \prod_\gamma J_\gamma(x) \qquad 13.7$$

**Definition:** *J is called a current operator* (its relationship to the electric current will be shown).

A number of particles participate in the interaction. As described by operators, the particles prior to interaction are annihilated and the particles present after interaction are created – a particle which is physically preserved is described as being annihilated and re-created. $H(x)$ can be represented as a Feynman node. Each line at the node corresponds to one particle in the interaction. In a single Feynman node there are no geometrical relationships with other matter, so it is not possible to say whether a particle's proper time is running forwards or backwards with respect to the reference frame clock. So a line for the annihilation of a particle, $\gamma$, may also represent the creation of the corresponding antiparticle $\bar{\gamma}$.

**Definition:** *Let* $\langle x, \alpha |$ *be the annihilation operator for a particle at* $(x, \alpha) = (x_0, \boldsymbol{x}, \alpha) \in T \otimes N$, *and let* $|\overline{x, \alpha}\rangle$ *be the creation operator for the antiparticle. Then the particle field* $\phi_\alpha(x): \mathscr{F} \to \mathscr{F}$ *is*

$$\phi_\alpha(x) = |\overline{x, \alpha}\rangle + \langle x, \alpha| \qquad 13.8$$

Then each line at the Feynman node corresponds to a particle field modelling the creation or annihilation of a particle. Clearly the hermitian conjugate of a particle field is the antiparticle field

$$\phi^\dagger{}_\alpha(x) = |\underline{x, \alpha}\rangle + \langle \overline{x, \alpha}| \qquad 13.9$$

In the general case $J_\gamma(x)$ is hermitian so it combines the particle and antiparticle fields

$$J_\gamma(x) = J_\gamma(\phi_\alpha(x), \phi^\dagger_\alpha(x)) \qquad 13.10$$

Then the general form of the interaction is

$$I_j(x_0) = :\sum_{x \in N} \prod_\gamma J_\gamma(|\overline{x,\alpha}\rangle + \langle \underline{x,\alpha}|, |\underline{x,\alpha}\rangle + \langle \overline{x,\alpha}|): \qquad 13.11$$

The colons reorder the creation and annihilation operators by placing all creation operators to the left of all annihilation operators, to ensure that false values are not generated corresponding to the annihilation of particles in the interaction in which they are created. Particular interactions can be postulated as operators with the general form of 13.11, we can examine whether the resulting theoretical properties correspond to the observed behaviour of matter.

**Definition:** *Let $\pi$ be the permutation such that $\tau_{\pi(n)} > \ldots \tau_{\pi(2)} > \tau_{\pi(1)}$ Then the time ordered product is*

$$T\{I(\tau_n)\ldots I(\tau_1)\} = I(\tau_{\pi(n)})\ldots I(\tau_{\pi(1)})$$

**Theorem:** *(Locality)*

$$\forall x, y \in T \otimes N \text{ such that } x - y \text{ is space-like } \langle [H(y), H(x)] \rangle = 0 \qquad 13.12$$

*Proof:* Iterate 7.2 from an initial condition at $t = 0$ given by $|f\rangle_0 \in \mathscr{F}$

$$|f\rangle_1 = \mu(1 - iI(0))|f\rangle_0$$

$$|f\rangle_2 = \mu^2(1 - iI(1))(1 - iI(0))|f\rangle_0$$

$$|f\rangle_3 = \mu^3(1 - iI(2))(1 - iI(1))(1 - iI(0))|f\rangle_0$$

Expand after $T$ iterations

$$|f\rangle_T = \mu^T \left( 1 + i \sum_{\tau_1 = 0}^{T-1} I(\tau) + (-i)^2 \sum_{\substack{\tau_2 = 0 \\ \tau_2 > \tau_1}}^{T-1} I(\tau_2) \sum_{\tau_1 = 0}^{T-1} I(\tau_1) + \ldots \right) |f\rangle_0 \qquad 13.13$$

Then 13.13 is

$$|f\rangle_T = \mu^T \left( 1 + \sum_{n=1}^{T} \frac{(-i)^n}{n!} \sum_{\substack{\tau_1 \ldots \tau_n = 0 \\ i \neq j \Rightarrow \tau_i \neq \tau_j}}^{T-1} T\{I(\tau_n)\ldots I(\tau_1)\} \right) |f\rangle_0 \qquad 13.14$$

There may be any number of particles in the initial state $|f\rangle_0 \in \mathscr{F}$ so 13.14 can be interpreted directly as a quantum logical statement meaning that, since an unknown number of interactions take place at unknown positions and unknown time, the final state is (named as) the weighted sum of the possibilities. Except for asymptotically free initial and final states, this statement ceases to make sense in the limit $T \to \infty$, which forces $N_S \to \mathbb{N}^3 \otimes S$ to ensure that particles remain in N. The expansion may reasonably be expected not to converge under such conditions, but there is no problem for bounded N and finite values of $T$ (i.e. stable fore and after states). By 13.6, 13.14 is

$$|f\rangle_T = \mu^T \left( 1 + \sum_{n=1}^{T} \frac{(-i)^n}{n!} \sum_{\substack{x^1 \ldots x^n \in T \otimes N_S \\ i \neq j \Rightarrow x_0^i \neq x_0^j}} T\{H(x^n)\ldots H(x^1)\} \right) |f\rangle_0 \qquad 13.15$$

Under Lorentz transformation of 13.15 the order of interactions, $H(x^i)$, can be changed in the time ordered product whenever $x^i - x^j$ is space-like. But this cannot affect the final state $|f\rangle_T$ for any $T \in \mathbb{N}$.

**Corollary:** *By 13.7 H factorises and the locality condition applies to the current operators.*

$$\forall x, y \in T \otimes N_S \text{ such that } x - y \text{ is space-like } \langle [J(y), J(x)] \rangle = 0 \qquad 13.16$$

## 14 Classical Law

**Theorem:** *In an inertial reference frame momentum is conserved.*

*Proof:* Classical momentum is the expectation of the momentum of a large number of particles. In the absence of interaction the expectation of momentum is constant for each particle by Newton's first law, 12.4. So it is sufficient to prove conservation of momentum in each particle interaction. Expand the interaction density, 13.11, as a sum of terms of the form

$$i(x_0) = \sum_{x \in N} h(x) = \sum_{x \in N} |\underline{x, \alpha}\rangle_1 \ldots |\underline{x, \alpha}\rangle_m \langle \underline{x, \alpha}|_{m+1} \ldots \langle \underline{x, \alpha}|_n \qquad 14.1$$

where $|\underline{x, \alpha}\rangle_i$ and $\langle \underline{x, \alpha}|_i$ are creation and annihilation operators for the particles and antiparticles in the interaction, given by 13.2. Suppress the spin indices by writing $\forall p \in M \ s = 1, 2, 3, 4 \ |p\rangle = |p, s\rangle$ and $|\underline{x}\rangle = |\underline{x, \alpha}\rangle$. We have from 14.1, $\forall n, m \in \mathbb{N}, n, m > 0$, $\forall$ plane wave $|p^1\rangle, \ldots, |p^n\rangle$

$$\langle p^1; \ldots; p^m | i(x_0) | p^{m+1}; \ldots; p^n \rangle = \langle p^1; \ldots; p^m | \sum_{x \in N} |\underline{x}\rangle^1 \ldots |\underline{x}\rangle^m \langle \underline{x}|^{m+1} \ldots \langle \underline{x}|^n | p^{m+1}; \ldots; p^n \rangle$$

Then, by 5.18

$$\langle p^1; \ldots; p^m | i(x_0) | p^{m+1}; \ldots; p^n \rangle = \sum_{x \in N} \sum_{\pi} \varepsilon(\pi) \prod_{i=1}^{m} \langle p^i | \underline{x} \rangle^{\pi(i)} \sum_{\pi'} \varepsilon(\pi') \prod_{j=m+1}^{n} \langle \underline{x}^j | p^{\pi'(j)} \rangle$$

which is a sum of terms of the form

$$\sum_{x \in N} \prod_{i=1}^{m} \langle q_i | \underline{x} \rangle_{\pi(i)} \prod_{j=m+1}^{n} \langle \underline{x}|_j p_{\pi'(j)} \rangle .$$

Using 13.4 and permuting $p_{\pi'(j)} \to p_j$ this reduces to a sum of terms of the form

$$\sum_{x \in N} \prod_{i=1}^{m} \langle q^i | \underline{\alpha} \rangle e^{iq^i \cdot x} \prod_{j=1}^{n} \langle \underline{\alpha} | p^j \rangle e^{-ip^j \cdot x} = \delta \left( \sum_{j=m+1}^{n} p^j - \sum_{i=1}^{m} q^i \right) \prod_{i=1}^{m} \langle q^i | \underline{\alpha} \rangle e^{-iq_0^i x_0} \prod_{j=1}^{n} \langle \underline{\alpha} | p_j \rangle e^{-p_0^j x_0}$$

by 3.12. Thus momentum is conserved in each particle interaction, and so is conserved universally by Newton's first law 12.4.

*Remark:* Conservation of momentum depends solely on the principle of homogeneity as expressed in 13.4,

and the mathematical properties of multiparticle vector space imposed by definition. Energy is not conserved in an individual interaction.

We are interested in changes in classical observable quantities. That is changes in the expectation, $\langle O \rangle$ of an observable, $O = O(x) = O(t, \mathbf{x})$. Since measurement is a combination of interactions, all observable quantities are composed of interaction operators, which, by 13.11, can be decomposed into fields. Thus physically observable discrete values are obtained from differentiable functions, and difference equations in the discrete quantities are obtained by integrating differential equations over one unit of time.

**Lemma:** *The equal time commutator between an observable operator O such that $O(x) = O(H(x))$ and the interaction density H obeys the commutation relation*

$$\forall \mathbf{x} \neq \mathbf{y}, [H(x), O(y)]_{x_0 = y_0} = 0 \qquad 14.2$$

*Proof:* Immediate from 13.16

**Theorem:** *The expectation of an observable operator $O(x) = O(H(x))$ obeys the differential equations*

$$\partial_0 \langle O(x) \rangle = i \langle [H(x), O(x)] \rangle + \langle \partial_0 O(x) \rangle$$
$$\text{For } \alpha = 1, 2, 3 \qquad \partial_\alpha \langle O(x) \rangle = \langle \partial_\alpha O(x) \rangle \qquad 14.3$$

*Proof:* By 7.2

$$\langle O(t+1) \rangle = \langle f |_t (1 + iI(t+1)) O(t+1) |\mu|^2 (1 - iI(t+1)) | f \rangle_t$$
$$= \langle f |_t i[I(t+1), O(t+1)] + O(t+1) | f \rangle_t$$

by 7.4, since the state is an eigenstate of $O$ and $|\mu|^2 = 1$. Then

$$\langle O(t+1) \rangle - \langle O(t) \rangle = \langle f |_{t+1} O(t+1) | f \rangle_{t+1} - \langle f |_t O(t) | f \rangle_t$$
$$= \langle f |_t i[I(t+1), O(t+1)] + O(t+1) | f \rangle_t - \langle f |_t O(t) | f \rangle_t$$

Using linearity of kets treated as operators and rearranging

$$\langle O(t+1) \rangle - \langle O(t) \rangle = i \langle [I(t+1), O(t+1)] \rangle + \langle O(t+1) - O(t) \rangle \qquad 14.4$$

The solution to 14.4 is the restriction to integer values of the solution of

$$\partial_0 \langle O(x) \rangle = i \langle [I(t), O(x)] \rangle + \langle \partial_0 O(x) \rangle \qquad 14.5$$

Using locality, 14.2, with $x_0 = y_0$ 14.5 is

$$\partial_0 \langle O(x) \rangle = i \langle \left[ \sum_{\mathbf{y} \in \mathbb{N}} H(x_0, \mathbf{y}), O(x) \right] \rangle + \langle \partial_0 O(x) \rangle \qquad 14.6$$

Using locality, 13.12, 14.6 reduces to the time-component of 14.3. The proof for $\alpha = 1, 2, 3$ is trivial.

**Corollary:** *Particles are point-like.*

**Note:** Position is only a numerical value derived from a configuration of matter in measurement, and it is not obvious that this requires that particles are themselves point-like.

*Proof:* By 14.2 and 14.3 changes in $\langle O(x) \rangle$ have no dependence on interactions except at the point $x$.

**Corollary:** *No observable particle effect may propagate faster than the speed of light.*

*Proof:* By 14.3 $O(x)$ has no space-like dependence on particle interactions for any space-like slice.

14.2 involves the commutation relation between the interaction density, $H$, and the observable, $O$. Since all physical processes are described by interactions, any observable operator is a combination of interaction operators, so observables are a combination of particle fields. Then 14.2 requires the commutators for particle fields. For fermions the creation operators anticommute, but commutation relations are obtained if the current, 13.10, is a composition of an even number of fermion fields.

## 15 The Photon Field

Photons are bosons, and having zero mass, the photon is its own antiparticle so that $\overline{|x, \alpha\rangle} = |x, \alpha\rangle$.

**Definition:** *By 13.8, the photon field is*

$$A_\alpha(x) = |\underline{x, \alpha}\rangle + \langle \underline{x, \alpha}| \qquad 15.1$$

which is hermitian, so only one photon field is necessary in the current, so $J = A$ is permissible and photons can be absorbed and emitted singly. The commutator is

$$[A_\alpha(x), A_\beta(y)] = [|\underline{x, \alpha}\rangle + \langle \underline{x, \alpha}|, |\underline{y, \beta}\rangle + \langle \underline{y, \beta}|] = \langle \underline{x, \alpha}|\underline{y, \beta}\rangle - \langle \underline{y, \beta}|\underline{x, \alpha}\rangle \qquad 15.2$$

Thus, by 12.10 and 13.4

$$[A_\alpha(x), A_\beta(y)] = \sum_r \eta(r) \int_M d^3p \langle \underline{\alpha}|p, r\rangle e^{-ip \cdot (x-y)} \langle p, r|\underline{\beta}\rangle - \langle \underline{\beta}|p, r\rangle e^{ip \cdot (x-y)} \langle p, r|\underline{\alpha}\rangle \qquad 15.3$$

The constraint that $A_\alpha(x)$ has only components of spin $\alpha$ is necessary if the interaction operator creates eigenstates of spin. This is observed; we assume that it also holds for time-like and longitudinal spin. Then $\langle \underline{\alpha}|p, r\rangle$ transforms as $w_\alpha(p, r)$ (defined in 11.1) under space inversion. So

$$\langle \underline{\beta}|-p, r\rangle\langle -p, r|\underline{\alpha}\rangle = \langle \underline{\alpha}|p, r\rangle\langle p, r|\underline{\beta}\rangle \qquad 15.4$$

since $w_\alpha(p, 0)$ has no space-like component and for $r = 1,2,3$ $w_\alpha(p, r)$ has no time like component. Now substitute $p \to -p$ in the second term of 15.3 at $x_0 = y_0$

$$[A(x), A(y)]_{x_0 = y_0} = 0 \qquad 15.5$$

Then by substituting $O = A$ in 14.2, and noting from 13.7 that the commutation relationship with the interaction density is determined by the commutation relationship with the current

$$\partial_\alpha \langle A_\beta(x) \rangle = \langle \partial_\alpha A_\beta(x) \rangle \qquad 15.6$$

The physical interpretation of 15.6 is that observable effects associated with photons depend only on changes in photon number; since photons can be absorbed or emitted singly the number of photons cannot be an eigenstate of an operator constructed from the interaction and cannot therefore be known. Let $\phi_\mu(x)$ be a gauge term, that is an arbitrary solution of

$$\partial_\mu \phi_\mu(x) = 0 \qquad 15.7$$

having no physical meaning. Then physical predictions from 15.6 are invariant under the gauge transformation $A(x) \to A(x) + \phi(x)$, and the value of $\langle A(x) \rangle$ is hidden by the gauge term. Differentiating 15.6 using 14.2 gives

$$\partial^2 \langle A(x) \rangle = \partial_\alpha \langle \partial_\alpha A(x) \rangle = i \langle [H(x), \partial_0 A(x)] \rangle + \langle \partial^2 A(x) \rangle \qquad 15.8$$

Differentiate twice and observe that $p^2 = 0$ for the photon so $\partial^2 |\underline{x, \alpha}\rangle = 0$. Then from 15.1

$$\partial^2 A(x) = 0 \qquad 15.9$$

Then 15.8 reduces to

$$\partial^2 \langle A(x) \rangle = i \langle [H(x), \partial_0 A(x)] \rangle \qquad 15.10$$

Given $H$, 15.10 can be calculated from the commutator between the fields

$$[\partial A_\alpha(x), A_\beta(y)] = \langle \partial x, \underline{\alpha} | y, \underline{\beta} \rangle - \langle y, \underline{\beta} | \partial x, \underline{\alpha} \rangle \qquad 15.11$$

But by 13.5 and 13.4

$$\langle \partial x, \underline{\alpha} | y, \underline{\beta} \rangle = -\sum_{r=0}^{3} \eta(r) \int_M d^3 p \langle \underline{\alpha} | p, r \rangle \langle p, r | \underline{\beta} \rangle i p e^{-ip \cdot (x-y)} \qquad 15.12$$

and

$$\langle y, \underline{\beta} | \partial x, \underline{\alpha} \rangle = \sum_{r=0}^{3} \eta(r) \int_M d^3 p \langle \underline{\beta} | p, r \rangle \langle p, r | \underline{\alpha} \rangle i p e^{ip \cdot (x-y)} \qquad 15.13$$

Substituting $p \to -p$ in 15.13 at $x_0 = y_0$ and using 15.4 and 15.11 gives, for the space-like components of the derivative

$$\text{For } i = 1, 2, 3, \ [\partial_i A(x), A(y)]_{x_0 = y_0} = 0$$

and for the time-like component

$$[\partial_0 A_\alpha(x), A_\beta(y)]_{x_0 = y_0} = -2i \sum_{r=0}^{3} \eta(r) \int_M d^3 p \langle \underline{\alpha} | p, r \rangle \langle p, r | \underline{\beta} \rangle p_0 e^{ip \cdot (x-y)} \qquad 15.14$$

**Theorem:** *The equal time commutator 15.14 satisfies locality, 13.16, if*

$$\langle \underline{\alpha} | p, r \rangle = \left(\frac{1}{2\pi}\right)^{\frac{3}{2}} \frac{w_\alpha(p, r)}{\sqrt{2 p_0}} \qquad 15.15$$

*Proof:* It follows from 15.15 that

$$\sum_{r=0}^{3} \eta(r) \langle \underline{\alpha} | p, r \rangle \langle p, r | \underline{\beta} \rangle = \frac{\eta(r) \delta_{\alpha\beta}}{16 \pi^3 p_0} \qquad 15.16$$

Substituting 15.16 into 15.14 shows locality is satisfied by the equal time commutator

$$[\partial_0 A(x), A(y)]_{x_0 = y_0} = -i g \delta_{xy} \qquad 15.17$$

Substituting 15.15 into 15.1 using 12.3 gives the photon field

$$A_\alpha(x) = \sum_{r=0}^{3} \eta(r) \int_M \frac{d^3 p}{\sqrt{2 p_0}} (e^{ip \cdot x} | p, r \rangle + e^{-ip \cdot x} \langle p, r |) w_\alpha(p, r) \qquad 15.18$$

By 15.16, 13.2 and 12.11

$$\langle x, \underline{\alpha} | y, \underline{\beta} \rangle = \frac{g_{\alpha\beta}}{8\pi^3} \int \frac{d^3 p}{2 p_0} e^{-ip \cdot (x-y)} \qquad 15.19$$

So the commutator, 15.2, is

$$[A_\alpha(x), A_\beta(y)] = \frac{g_{\alpha\beta}}{8\pi^3} \int_M \frac{d^3 p}{2 p_0} (e^{-ip \cdot (x-y)} - e^{ip \cdot (x-y)}) \qquad 15.20$$

**Theorem:** *15.20 is zero outside the light cone.*

*Proof:* The proof follows the text books, e.g. [1], and is left to the reader.

**Theorem:** $\langle A(x) \rangle$ satisfies the Lorentz gauge condition

$$\partial_\alpha \langle A_\alpha(x) \rangle = 0 \qquad 15.21$$

*Proof:* by 15.6

$$\partial_\alpha \langle A_\alpha(x) \rangle = \langle \partial_\alpha A_\alpha(x) \rangle$$

$$= \langle \sum_{r=0}^{3} \eta(r) \int_M \frac{d^3 p}{\sqrt{2p_0}} (e^{ip \cdot x} | p, r \rangle + e^{-ip \cdot x} \langle p, r |) i(p_\alpha - p_\alpha) w_\alpha(p, r) \rangle$$

by differentiating 15.18. But this is zero which establishes 15.21.

## 16 The Dirac Field

**Definition:** *By 13.8, the Dirac field is*

$$\psi_\alpha(x) = \overline{|x, \alpha\rangle} + \langle x, \alpha| \qquad 16.1$$

We know from observation that a Dirac particle can be an eigenstate of position. Any physical configuration can only be a combination of particle interactions so it is possible to form the position operator

$$Z(X) = \sum_{x \in X} |x\rangle\langle x| \qquad 16.2$$

from the current 13.10, for any region $X$ which can be as small as the apparatus will allow. Position kets are a basis, so 16.2 reduces to

$$Z(x) = |x\rangle\langle x|$$

up to the resolution of the apparatus. Current can only generate eigenstates of spin and position if it does not mix basis states, so

$$\forall x \in N \ |\underline{x}, \alpha\rangle = |x, \alpha\rangle \qquad 16.3$$

Then by 12.1

$$\langle \underline{\alpha} | p, r \rangle = \left(\frac{1}{2\pi}\right)^{\frac{3}{2}} u_\alpha(p, r) \qquad 16.4$$

and by 18.8

$$\langle \underline{x}, \alpha | = \left(\frac{1}{2\pi}\right)^{\frac{3}{2}} \sum_r \int_M d^3 p \, u_\alpha(p, r) e^{-ip \cdot x} \langle p, r | \qquad 16.5$$

**Definition:** *The Dirac adjoint of the annihilation operator $\langle \underline{x}, \alpha |$ is*

$$|\underline{x}, \hat{\alpha}\rangle = \sum_\mu |\underline{x}, \mu\rangle \gamma^0_{\mu\alpha} = \left(\frac{1}{2\pi}\right)^{\frac{3}{2}} \sum_r \int_M d^3 p \, \hat{u}_\alpha(p, r) e^{ip \cdot x} | p, r \rangle \qquad 16.6$$

Similarly by 12.2

$$\langle \bar{\alpha} | p, r \rangle = \left(\frac{1}{2\pi}\right)^{\frac{3}{2}} \bar{v}_\alpha(p, r) \qquad 16.7$$

and by 18.8

$$\overline{|x, \alpha\rangle} = \left(\frac{1}{2\pi}\right)^{\frac{3}{2}} \sum_r \int_M d^3p\, v_\alpha(p, r) e^{ip \cdot x} | p, r \rangle \qquad 16.8$$

**Definition:** *The Dirac adjoint of the creation operator $|x, \alpha\rangle$ is*

$$\langle \overline{x, \hat{\alpha}} | = \sum_\mu \langle \overline{x, \mu} | \gamma^0_{\mu\alpha} = \left(\frac{1}{2\pi}\right)^{\frac{3}{2}} \sum_r \int_M d^3p\, \hat{v}_\alpha(p, r) e^{ip \cdot x} | p, r \rangle \qquad 16.9$$

**Definition:** *The Dirac adjoint of the field is*

$$\hat{\bar{\psi}}_\alpha(x) = \psi^\dagger_\mu(x) \gamma^0_{\mu\alpha} = \overline{|x, \hat{\alpha}\rangle} + \langle \overline{x, \hat{\alpha}} | \qquad 16.10$$

**Theorem:** *The anticommutation relations for the Dirac field and Dirac adjoint and obey*

$$\{\psi_\nu(x), \psi_\lambda(y)\} = \{\hat{\bar{\psi}}_\mu(x), \hat{\bar{\psi}}_\kappa(y)\} = 0 \qquad 16.11$$

$$\{\psi_\alpha(x), \hat{\bar{\psi}}_\beta(y)\}_{x_0 = y_0} = \gamma^0_{\alpha\beta} \delta_{xy} \qquad 16.12$$

*Proof:* 16.11 follows from the definitions, 16.1 and 16.10 By 6.10 and 5.16 we have

$$\{\psi_\alpha(x), \hat{\bar{\psi}}_\beta(y)\} = \{\langle x, \alpha|, |y, \hat{\beta}\rangle\} + \{\overline{|x, \alpha\rangle}, \langle \overline{y, \hat{\beta}} |\}$$

$$= \langle x, \alpha | y, \hat{\beta} \rangle + \langle \overline{y, \hat{\beta}} | \overline{x, \alpha} \rangle^T \qquad 16.13$$

where $^T$ denotes that $\alpha$ and $\beta$ are transposed. By 16.5 and 16.6, and using 12.11.

$$\langle x, \alpha | y, \hat{\beta} \rangle = \frac{1}{8\pi^3} \sum_r \int_M d^3p\, u_\alpha(p, r) \hat{u}_\beta(p, r) e^{-ip \cdot (x-y)}$$

$$= \frac{1}{8\pi^3} \int_M \frac{d^3p}{2p_0} (p \cdot \gamma + m)_{\alpha\beta} e^{-ip \cdot (x-y)} \qquad 16.14$$

by 10.7. Likewise for the antiparticle, by 16.8 and 16.9

$$\langle \overline{y, \hat{\beta}} | \overline{x, \alpha} \rangle^T = \frac{1}{8\pi^3} \sum_r \int_M d^3p\, v_\alpha(p, r) \hat{v}_\beta(p, r)\, e^{ip \cdot y - ix \cdot p}$$

$$= \frac{1}{8\pi^3} \int_M \frac{d^3p}{2p_0} (p \cdot \gamma - m)_{\alpha\beta}\, e^{ip \cdot (x-y)} \qquad 16.15$$

by 10.12. Substituting $p \to -p$ at $x_0 = y_0$ in 16.15 gives

$$\langle \overline{y, \hat{\beta}} | \overline{x, \alpha} \rangle_{x_0 = y_0} = \frac{1}{8\pi^3} \int_M \frac{d^3p}{2p_0} (2p_0 \gamma^0 - p \cdot \gamma - m) e^{-ip \cdot (x-y)} \qquad 16.16$$

So, by 16.13, adding 16.14 and 16.16 at $x_0 = y_0$ gives the equal time anticommutator

$$\{\psi_\alpha(x), \hat{\bar{\psi}}_\beta(y)\}_{x_0 = y_0} = \frac{1}{8\pi^3} \gamma^0_{\alpha\beta} \int_M d^3p\; e^{-ip \cdot (x-y)} \qquad 16.17$$

16.12 follows immediately.

**Theorem:** *The anticommutation relations for the Dirac field and the Dirac adjoint obeys locality, 13.16.*

*Proof:* By 16.14

$$\langle \underline{x, \alpha} | \underline{y, \hat{\beta}} \rangle = \frac{1}{8\pi^3}(i\partial \cdot \gamma + m) \int_M \frac{d^3p}{2p_0} e^{-ip \cdot (x-y)} \quad \quad 16.18$$

And by 16.15

$$\overline{\langle y, \hat{\beta} | x, \alpha \rangle}^T = -\frac{1}{8\pi^3}(i\partial \cdot \gamma + m) \int_M \frac{d^3p}{2p_0} e^{ip \cdot (x-y)} \quad \quad 16.19$$

By 16.13 the anticommutator is found by adding 16.18 and 16.19

$$\{\psi_\alpha(x), \hat{\psi}_\beta(y)\} = \frac{1}{8\pi^3}(i\partial \cdot \gamma + m) \int_M \frac{d^3p}{2p_0}(e^{-ip \cdot (x-y)} - e^{ip \cdot (x-y)}) \quad \quad 16.20$$

16.20 is and zero outside the light cone. The proof follows the text books, e.g. [1], and is left to the reader.

## 17 The Non-Perturbative Solution

Because local phase is a freedom in the definition of Hilbert space and can give no physical results we have that $U(1)$ local gauge symmetry is preserved in interaction (section 2). Then we have the intuitively appealing minimal interaction characterised by the emission or absorption of a photon by a Dirac particle. According to 13.7 an interaction $H$ between photons and Dirac particles is described by a combination of particle currents, which, by 13.10, are themselves hermitian combinations of particle fields.

**Definition:** *The photon current operator is $A(x)$*

**Definition:** *The Dirac current operator is*

$$j_\alpha(x) = :\hat{\psi}_\mu(x)\gamma^\alpha_{\mu\nu}\psi_\nu(x): = :\hat{\psi}(x)\gamma^\alpha\psi(x): \quad \quad 17.1$$

**Postulate:** *Let e be the electromagnetic coupling constant. The electromagnetic interaction density is*

$$H(x) = ej(x) \cdot A(x) = e:\hat{\psi}(x)\gamma \cdot A(x)\psi(x): \quad \quad 17.2$$

**Lemma:**

$$\langle \partial \cdot j(x) \rangle = 0 \quad \quad 17.3$$

*Proof:* Using the definitions 16.1 and 16.10 to expand 17.1

$$j_\alpha(x) = |\underline{x, \hat{\mu}}\rangle\gamma^\alpha_{\mu\nu}|\overline{x, \nu}\rangle + |\underline{x, \hat{\mu}}\rangle\gamma^\alpha_{\mu\nu}\langle\underline{x, \nu}| - \gamma^\alpha_{\mu\nu}|\overline{x, \nu}\rangle\langle\overline{x, \hat{\mu}}| + \langle\overline{x, \hat{\mu}}|\gamma^\alpha_{\mu\nu}\langle\underline{x, \nu}| \quad \quad 17.4$$

where the summation convention is used for the repeated indices, $\mu$ and $\nu$. In classical situations we only consider states of a definite number of Dirac particles, so the expectation of the pair creation and annihilation terms is zero by 4.1. Using 16.5 and 16.6 and differentiating the particle term in 17.4

$$\partial_\alpha |\underline{x, \hat{\mu}}\rangle\gamma^\alpha_{\mu\nu}\langle\underline{x, \nu}| = \frac{1}{8\pi^3}\sum_{r,s}\int_M d^3p \int_M d^3q \, i\hat{u}(p, r)(q \cdot \gamma - p \cdot \gamma)u(q, s)e^{ix \cdot (q-p)}|p, r\rangle\langle q, s|$$

Using 16.8 and 16.9 and differentiating the antiparticle term in 17.4

$$\partial_\alpha \gamma^\alpha_{\mu\nu} \overline{|x,\nu\rangle\langle x,\hat{\mu}|} = \frac{1}{8\pi^3} \sum_{r,s} \int_M d^3p \int_M d^3q\, i\hat{v}(q,r)(p\cdot\gamma - q\cdot\gamma)v(p,s)e^{ix\cdot(p-q)}|p,r\rangle\langle q,s|$$

Here $v$ and $\hat{v}$ have been ordered so that the spin index can be unambiguously omitted. 17.3 follows by differentiating 17.4 and using 10.6 and 10.11.

**Lemma:**
$$[j_0(x), j_\alpha(x)] = 0 \qquad 17.5$$

*Proof:* $\quad [\psi(x), j_\alpha(x)] = [\psi(x), :\hat{\psi}(x)\gamma^\alpha\psi(x):] = \{\psi(x), \hat{\psi}(x)\}\gamma^\alpha\psi(x)$

$$= \gamma^0\gamma^\alpha\psi(x) \qquad 17.6$$

by 16.12. Take the hermitian conjugate and apply 10.5

$$[j_\alpha(x), \psi^\dagger(x)] = \psi^\dagger(x)\gamma^{\alpha\dagger}\gamma^0 = \hat{\psi}(x)\gamma^\alpha$$

Postmultiply by $\gamma^0$

$$[j_\alpha(x), \hat{\psi}(x)] = \hat{\psi}_\mu(x)\gamma^\alpha\gamma^0 \qquad 17.7$$

So, by commuting the terms,

$$[j_0(x), j_\alpha(x)] = [:\hat{\psi}(x)\gamma^0\psi(x):, j_\alpha(x)]$$

$$= \hat{\psi}(x)\gamma^0[\psi(x), j_\alpha(x)] + [\hat{\psi}(x), j_\alpha(x)]\gamma^0\psi(x)$$

$$= \hat{\psi}(x)\gamma^0\gamma^0\gamma^\alpha\psi(x) - \hat{\psi}(x)\gamma^\alpha\gamma^0\gamma^0\psi(x)$$

using 17.6 and 17.7. 17.5 follows from 10.5

**Theorem:** $\langle j \rangle$ *is a classical conserved current, i.e.*

$$\partial \cdot \langle j(x) \rangle = 0 \qquad 17.8$$

*Proof:* Substituting $O = j_\alpha$ in 14.3

$$\partial_a \langle j_\alpha(x)\rangle = i\langle[H(x), j_0(x)]\rangle + \langle\partial_\alpha j_\alpha(x)\rangle$$

17.8 follows from 17.3 and 17.5, so $\langle j \rangle$ is conserved.

**Theorem:** $\langle j_0 \rangle$ *can be identified with classical electric charge density, i.e.*

$$\forall |f\rangle \in \mathscr{F},\ \langle j_0(x)\rangle = |\langle x|f\rangle|^2 - |\langle f|\overline{x}\rangle|^2 \qquad 17.9$$

*Proof:* It is straightforward from 6.2 that $j$ is additive for multiparticle states, so it is sufficient to show the theorem for a one particle state $|f\rangle \in \mathbb{H}$. By 17.4

$$\langle j_0(x)\rangle = \langle f|\underline{x},\hat{\mu}\rangle\gamma^0_{\mu\nu}\langle\underline{x},\nu|f\rangle - \gamma^0_{\mu\nu}\langle f|\overline{x,\nu}\rangle\langle\overline{x,\hat{\mu}}|f\rangle$$

$$= \langle f|\underline{x}\rangle\gamma^0\gamma^0\langle\underline{x}|f\rangle - \langle\overline{x}|f\rangle\gamma^0\gamma^0\langle f|\overline{x}\rangle$$

by ordering terms so that the spinor indices can be suppressed. Then 16.14 follows from 16.3 and 10.5

Except in so far as $\mathbb{A}2$ was used to justify an analysis of measurement, classical law does not form part of the assumptions, and according to $\mathbb{L}7$, the claim that the minimal interaction is the cause of the electromagnetic force requires:

**Theorem:** $\langle A(x) \rangle$ *satisfies Maxwell's Equations*

$$\partial^2 \langle A_\alpha(x) \rangle - \partial_\alpha \partial_\mu \langle A_\mu(x) \rangle = -e \langle j_\alpha(x) \rangle \qquad 17.10$$

**Corollary:** *Maxwell's equations simplify immediately to their form in the Lorentz gauge*

$$\partial^2 \langle A(x) \rangle = -e \langle j(x) \rangle \qquad 17.11$$

*Proof:* By 15.21 it is sufficient to prove the corollary. By 15.10 and 17.2

$$\partial^2 \langle A(x) \rangle = i \langle [j(x) \cdot A(x), \partial_0 A(x)] \rangle$$

17.11 follows immediately from 15.17.

**Theorem:** *(Classical gauge invariance). Let g be an arbitrary differentiable function. Then classical properties are invariant under gauge transformation of the photon field given by*

$$\langle A_\alpha(x) \rangle \to \langle A_\alpha(x) + \partial_\alpha g(x) \rangle = \langle A_\alpha(x) \rangle + \partial_\alpha g(x) \qquad 17.12$$

*Proof:* It is a well known result following from 17.10 that the classical properties of the electromagnetic field depend only on derivatives of $\langle A(x) \rangle$, defined by

$$F_{\alpha\beta} \equiv \partial_\alpha \langle A_\beta(x) \rangle - \partial_\beta \langle A_\alpha(x) \rangle$$

Then $F_{\alpha\beta}$ is clearly invariant under 17.12. Although classical electrodynamics is gauge invariant, the Lorentz gauge, 15.21, is here theoretically determined and we have $\partial_\alpha g = 0$.

## 18  Feynman Rules

**Definition:** *For any vector p, such that $p^2 = m^2$, let $\tilde{p} = (\tilde{p}_0, \boldsymbol{p})$ be a matrix for any $\tilde{p}_0 \in \mathbb{R}$. $\tilde{p}$ satisfies the identity*

$$\tilde{p}_0^2 - p_0^2 \equiv \tilde{p}^2 - m^2 \qquad 18.1$$

**Lemma:** *For $x > 0$, $\varepsilon > 0$ we have the identities*

$$\frac{e^{i(p_0 - i\varepsilon)x}}{2(p_0 - i\varepsilon)} \equiv \frac{-i}{2\pi} \int_{-\infty}^{\infty} d\tilde{p}_0 \frac{e^{-i\tilde{p}_0 x}}{\tilde{p}_0^2 - (p_0 - i\varepsilon)^2} \equiv \frac{-i}{2\pi} \int_{-\infty}^{\infty} d\tilde{p}_0 \frac{e^{-i\tilde{p}_0 x}}{\tilde{p}^2 - m^2 + 2ip_0\varepsilon + \varepsilon^2} \qquad 18.2$$

$$\frac{e^{i(p_0 - i\varepsilon)x}}{2} \equiv \frac{-i}{2\pi} \int_{-\infty}^{\infty} d\tilde{p}_0 \frac{\tilde{p}_0 e^{-i\tilde{p}_0 x}}{\tilde{p}^2 - m^2 + 2ip_0\varepsilon + \varepsilon^2} \qquad 18.3$$

and for $x < 0$  $\varepsilon > 0$ we have the identities

$$\frac{e^{-i(p_0 - i\varepsilon)x}}{2(p_0 - i\varepsilon)} \equiv \frac{-i}{2\pi} \int_{-\infty}^{\infty} d\tilde{p}_0 \frac{e^{-i\tilde{p}_0 x}}{\tilde{p}^2 - m^2 + 2ip_0\varepsilon + \varepsilon^2} \qquad 18.4$$

$$\frac{e^{-i(p_0 - i\varepsilon)x}}{2} \equiv \frac{-i}{2\pi} \int_{-\infty}^{\infty} d\tilde{p}_0 \frac{\tilde{p}_0 e^{-i\tilde{p}_0 x}}{\tilde{p}^2 - m^2 + 2ip_0\varepsilon + \varepsilon^2} \qquad 18.5$$

*Proof:* These are evaluated as contour integrals and the proofs are left to the reader.

**Definition:** *The step function is given $\forall x \in \mathbb{R}$, by*

$$\Theta(x) = \begin{cases} 0 & \text{if } x \leq 0 \\ 1 & \text{if } x > 0 \end{cases}$$

Let $|g\rangle \in \mathcal{F}$ be a measured state at time T. $\langle g|f\rangle_T$ can be evaluated iteratively from 13.15 by using 6.5. The result is the sum of the terms generated by the braket between $\langle x^n, \alpha|$ and every earlier creation operator $|x^j, \alpha\rangle$ and every particle in $|f\rangle_0$, and the braket between $|x^n, \alpha\rangle$ and every later annihilation operator $\langle x^j, \alpha|$ and every particle in the final state $\langle g|$ (all other brakets are zero). This procedure is repeated for every creation and annihilation operator in 17.4, and for every term in 13.15. To keep check on the brakets so formed, each factor $I_j(x_0)$ in 13.11 is represented as a Feynman node. Each line at the node corresponds to one of the particles in the interaction and to one of the particle fields $\overline{|x, \alpha\rangle} + \langle x, \alpha|$ in $I_j(x_0)$. Then when the braket is formed the corresponding connection between the nodes is made in a diagram. Each internal connecting line, or propagator, is associated with a particular particle type. Photons are denoted by wavy lines, and Dirac particles by arrowed lines, so that for particles the arrow is in the direction of time ordering in 13.15, and for antiparticles the arrow is opposed to the time ordering. In this way all time ordered diagrams are formed by making each possible connection, from the creation of a particle to the annihilation of a particle of the same type, and we calculate rules to evaluate the diagram from 13.15. There is an overall factor $1/n!$ for a diagram with $n$ vertices. The vertices, $x^n$, are such that $n \neq j \Rightarrow x_0^n \neq x_0^j$ and, by examination of 13.15 and 17.2, generate the expression

$$e \sum_{x^n \in T \otimes N} -i\gamma \qquad 18.6$$

The initial and final states are expressed as plane wave expansions so that the time invariant inner product, 12.10, can be used so that $\mu$ which appears in 13.15 is set to 1. Plane waves span $\mathcal{F}$, so without loss of generality can be used for the initial and final states. Then for an initial particle the state $|p, r\rangle$ connected to the node $x^n$ gives, from 13.4

$$\langle \underline{x^n, \alpha} |p, r\rangle = \langle \underline{\alpha} |p, r\rangle e^{-ip \cdot x^n}$$

So for a photon, by 15.15

$$\langle \underline{x^n, \alpha} |p, r\rangle = \left(\frac{1}{2\pi}\right)^{\frac{3}{2}} \frac{w_\alpha(p, r)}{\sqrt{2p_0}} e^{-ip \cdot x^n} \qquad 18.7$$

for a Dirac particle, by 16.4

$$\langle \underline{x^n, \alpha} |p, r\rangle = \left(\frac{1}{2\pi}\right)^{\frac{3}{2}} u_\alpha(p, r) e^{-ip \cdot x^n}$$

and for an antiparticle, by 16.7

$$\langle \overline{x^n, \hat{\alpha}} | p, r \rangle = \left(\frac{1}{2\pi}\right)^{\frac{3}{2}} \hat{v}_\alpha(p, r) e^{-ip \cdot x^n} \qquad 18.8$$

Similarly for final particles in the state $\langle p, r |$ connected to the node $x^n$ gives for a photon

$$\left(\frac{1}{2\pi}\right)^{\frac{3}{2}} \frac{w_\alpha(p, r)}{\sqrt{2p_0}} e^{ip \cdot x^n}$$

for a Dirac particle

$$\left(\frac{1}{2\pi}\right)^{\frac{3}{2}} \hat{u}_\alpha(p, r) e^{ip \cdot x^n}$$

and for an antiparticle

$$\left(\frac{1}{2\pi}\right)^{\frac{3}{2}} v_\alpha(p, r) e^{ip \cdot x^n} \qquad 18.9$$

The time ordered product in 13.15 leads to an expression for the photon propagator

$$\Theta(x_0^n - x_0^j)\langle \underline{x^n, \alpha} | \underline{x^j, \beta} \rangle + \Theta(x_0^j - x_0^n)\langle \underline{x^j, \beta} | \underline{x^n, \alpha} \rangle^T \qquad 18.10$$

18.10 is to be compared with the decomposition of distributions into advanced and retarded parts according to the method of Epstein and Glaser which also excludes $x_0^n = x_0^j$ [17][5]. Indeed our analysis of the origin of the ultraviolet divergence is essentially the same as that given by Scharf [17]. The difference between this treatment and Scharf is that our limiting procedure uses a discrete space and here the "switching off and switching on" of the interaction at $x_0^n = x_0^j$ is a physical constraint meaning that only one interaction takes place for each particle in any instant, as discussed in section 7. By 15.19 18.10 is

$$\frac{g_{\alpha\beta}}{8\pi^3} \int_M \frac{d^3p}{2p_0} [\Theta(x_0^n - x_0^j) e^{-ip \cdot (x^n - x^j)} + \Theta(x_0^j - x_0^n) e^{ip \cdot (x^n - x^j)}]$$

Use 18.2 in the first term, recalling that $m^2 = 0$, and use 18.4 and substitute $p \to -p$ in the second term. Then we have

$$-i \frac{g_{\alpha\beta}}{16\pi^4} \int_M \frac{d^3p}{2p_0 \varepsilon} \lim_{\to 0^+} \int_{-\infty}^\infty d\tilde{p}_0 [\Theta(x_0^n - x_0^j) + \Theta(x_0^j - x_0^n)] \frac{e^{-i\tilde{p} \cdot (x^n - x^j)}}{\tilde{p}^2 + 2ip_0\varepsilon + \varepsilon^2} \qquad 18.11$$

For each node the Dirac current generates two propagators, one for the field and one for the adjoint. The field either annihilates a particle or creates an antiparticle, and is represented by an arrowed line pointing towards the vertex. The field $\psi_\alpha(x^n)$ at vertex $n$ acting on vertex $j$, generates the propagator arrowed from $j$ to $n$

$$\Theta(x_0^n - x_0^j)\langle \underline{x^n, \alpha} | \underline{x^j, \hat{\beta}} \rangle - \Theta(x_0^j - x_0^n)\langle \overline{x^j, \hat{\beta}} | \overline{x^n, \alpha} \rangle^T \qquad 18.12$$

The Dirac adjoint field creates a particle or annihilates an antiparticle, and is represented by an arrowed line pointing away from the vertex. The adjoint $\hat{\psi}_\alpha(x^n)$ generates the propagator arrowed from $n$ to $j$

$$\Theta(x_0^n - x_0^j)\langle \overline{x^n, \hat{\alpha}} | \overline{x^j, \beta} \rangle - \Theta(x_0^j - x_0^n)\langle \underline{x^j, \beta} | \underline{x^n, \hat{\alpha}} \rangle^T \qquad 18.13$$

The time ordered product in 13.15 is unaffected under the interchange of $(x^n, \alpha)$ and $(x^j, \beta)$. By interchanging $(x^n, \alpha)$ and $(x^j, \beta)$ in the diagram, we find for the adjoint propagator arrowed from $j$ to $n$

$$\Theta(x_0^j - x_0^n)\overline{\langle x^j, \hat{\beta}|x^n, \alpha\rangle}^{\mathrm{T}} + \Theta(x_0^n - x_0^j)\underline{\langle x^n, \alpha|x^j, \hat{\beta}\rangle} \qquad 18.14$$

18.14 is identical to 18.12, the expression for the Dirac propagator arrowed from $j$ to $n$, so we do not distinguish whether an arrowed line in a diagram is generated by the field or the adjoint field. Similarly we find that the photon propagator, 18.10 is unchanged under interchange of the nodes, so we identify all diagrams which are the same apart from the ordering of the vertices and remove the overall factor $1/n!$ for a diagram with $n$ vertices. By 16.14 and 16.15, 18.12 is

$$\frac{\Theta(x_0^n - x_0^j)}{8\pi^3}\int_M \frac{d^3\boldsymbol{p}}{2p_0}(ip\cdot\gamma + m)e^{-ip\cdot(x^n - x^j)} + \frac{\Theta(x_0^j - x_0^n)}{8\pi^3}\int_M \frac{d^3\boldsymbol{p}}{2p_0}(ip\cdot\gamma - m)e^{ip\cdot(x^n - x^j)} \qquad 18.15$$

Use 18.2 and 18.3 in the first term, and use 18.4 and 18.5 and substitute $\boldsymbol{p}\to -\boldsymbol{p}$ in the second term. Then the propagator 18.15 is

$$-i\frac{g_{\alpha\beta}}{16\pi^4}\int_M \frac{d^3\boldsymbol{p}}{2p_0\varepsilon} \mathop{\mathrm{Lim}}_{\varepsilon\to 0^+}\int_{-\infty}^{\infty} d\tilde{p}_0[\Theta(x_0^n - x_0^j) + \Theta(x_0^j - x_0^n)]\frac{(ip\cdot\gamma + m)e^{-i\tilde{p}\cdot(x^n - x^j)}}{\tilde{p}^2 - m^2 + 2ip_0\varepsilon + \varepsilon^2}$$

Now collect all the exponential terms with $x^n$ in the exponent under the sum 18.6, and observe that the sum over space is a momentum conserving delta function by 3.12. Then integrate over momentum space and impose conservation of momentum at each vertex, leaving for each independent internal loop

$$\frac{1}{8\pi^3}\int_M \frac{d^3\boldsymbol{p}}{2p_0} \qquad 18.16$$

Only the time component remains in the exponents for the external lines 18.7 - 18.9. Introduce a finite cutoff $\Lambda\in\mathbb{N}$ by writing the improper integral

$$\frac{-i}{2\pi}\int_{-\infty}^{\infty} d\tilde{p}_0 = \mathop{\mathrm{Lim}}_{\Lambda\to\infty}\int_{-\Lambda\pi}^{\Lambda\pi} d\tilde{p}_0$$

and instructing that the limits $\Lambda\to\infty$, $\varepsilon\to 0^+$ should be taken after calculation of all formulae. Then the photon propagator, 18.11 reduces to

$$-\frac{ig_{\alpha\beta}}{2\pi}\int_{-\Lambda\pi}^{\Lambda\pi} d\tilde{p}_0 \frac{(1 - \delta_{x_0^n x_0^j})e^{i\tilde{p}_0(x_0^n - x_0^j)}}{\tilde{p}^2 + 2ip_0\varepsilon + \varepsilon^2} \qquad 18.17$$

For a Dirac particle, $p_0 > 0$, so we can also simplify the denominator by shifting the pole under the limit $\varepsilon\to 0^+$. Thus the Dirac propagator arrowed from $j$ to $n$ is

$$\frac{-i}{2\pi}\int_{-\Lambda\pi}^{\Lambda\pi} d\tilde{p}_0 \frac{(1 - \delta_{x_0^n x_0^j})(\tilde{p}\cdot\gamma + m)_{\alpha\beta}e^{-i(x_0^n - x_0^j)\tilde{p}_0}}{\tilde{p}^2 - m^2 + i\varepsilon} \qquad 18.18$$

The propagators, 18.17 and 18.18, vanish for $x_0^j = x_0^n$, and are finite in the limit $\Lambda\to\infty$, since the integrands oscillate and tend to zero as $p_0\to\infty$. Loop integrals (18.16) are proper and the denominators do not vanish so the ultraviolet divergence and the infrared catastrophe are absent, provided that the limits $\Lambda\to\infty$ and $\varepsilon\to 0^+$ are not taken prematurely. In the denominator of 18.17, $\varepsilon^2$ plays the role of the small photon mass commonly used to treat the infrared catastrophe, and as with Scharf's treatment there is no additional requirement to include a photon mass. The standard rules are obtained by ignoring $\delta_{x_0^n x_0^j}$ in the numerator of 18.17 and 18.18, and observing that for $\Lambda \in \mathbb{N}$, the sums over time in the vertex 18.6 act as $\tilde{p}_0$ conserving $\delta$ functions ($\tilde{p}_0$ is not energy since energy was defined in section 8 to be on mass shell, but $\tilde{p}_0$ is equal to the energy of the measured, initial and final, states).

Thus the discrete theory modifies the standard rules for the propagators and justifies the subtraction of divergent quantities, but here this is no ad hoc procedure but regularisation by the subtraction of a term which recognises that a particle cannot be annihilated at the instant of its creation. To see that this subtraction is effectively the same as that described by Scharf we replace $\Theta$ in the propagators 18.10 and 18.12 with a monotonous $C^\infty$ function $\chi_0$ over $\mathbb{R}^1$ with

$$\chi_0(t) = \begin{cases} 0 & \text{for } t \leq 0 \\ 1 & \text{for } t \geq \chi \end{cases}$$

Then we observe that in the limit as the discrete unit of time $\chi \to 0$ the sums over space become integrals since these are well defined [17]. In the limit M is replaced by $\mathbb{R}^3$ and 18.10 and 18.12 are replaced by distributions which have been split with causal support. When the distributions are combined in 18.17 and 18.18 we obtain the usual Feynman rules, together with a term coming from $\delta_{x_0^n x_0^j}$, which gives a distribution with point support in the limit (c.f. Scharf [17] 3.2.46). The most straightforward way to determine the effect of this term is to consider the non-perturbative solution (section 17). This allows us to impose three regularisation conditions on the propagator, that it is independent of lattice spacing $\chi$ at low energies and the renormalised mass and charge adopt their bare values, since bare mass and charge appear in Maxwell's equations.

This in no way contradicts the calculation that the apparent or "running" coupling constant exhibits an energy dependency in scattering due to perturbative corrections. But it shows that this dependency is removed in the calculation of the expectation of the current, and enables us to regularise the theory to the low energy value. The calculation of effective charge [15].(7.96) by the summation of one particle irreducible insertions [15].(7.94) into the photon propagator breaks down to any finite order, so the limit may not be taken. More generally the renormalisation group arguments leading to the behaviour of the Callan-Symanzik equation depend upon the sum of a geometric series [15].(10.27) which does not converge at the Landau pole. The Landau pole is absent in a model in which there is a fundamental minimum unit of time since high energies correspond to short interaction times and in a discrete model this implies fewer interactions. In the limit as the discrete unit of time goes to zero the pole suggests nothing more serious than the failure of an iterative solution, not the failure of the model.

## 19  Yang-Mills Fields and Quark Confinement

Following the arguments of section 9 we seek to replace the discrete time evolution equation 7.2 with a covariant equation of motion, which should have the form of 9.2, but with the addition of an interaction term

$$(i\partial \cdot \Gamma + H(x))f = mf \qquad \qquad 19.1$$

Now we seek to generalise to the case where we do not have a simple Dirac particle, but a composition of Dirac particles $|x_i, q_i\mu_i\rangle$ where $i$ is an index (colour), and $q_i$ is a flavour of quark. There are two straight-forward ways of forming states which are symmetrical in the colour index. The usual version of SU(3) is formed by creating linear combinations of quark states $|x_i, q_i\mu_i\rangle$, but it is interesting also to consider a model built from multiplets of quarks. Such models are generally rejected as being non-renormalisable, but deeper analysis will show that this is not so because the quarks are decoupled within the multiplet, and that the model has similar properties to the standard model.

Whether quarks are confined by the physical interactions described here, or whether there is some deeper confining mechanism in qcd, the current treatment provides a model with the principle features of the strong interactions, and is applicable whenever substructure is not relevant. A general state of three quarks in $\mathcal{H}$, as defined in 4.3

$$|f\rangle = (|x_1, q_1, \mu_1\rangle, |x_2, q_2, \mu_2\rangle, |x_3, q_3, \mu_3\rangle) \qquad 19.2$$

has a co-ordinate $x_i$ for the position of each quark. But the physical space is the subspace of $\mathcal{H}$ generated from 19.1. So physical states are either directly created from the creation operators appearing in $H(x)$, or evolve from them from the corresponding non-interacting a wave equation.

The spin statistics theorem forbids even multiplets but it is possible to build a model in which the field operators create or annihilate triplets of quarks. There is no way physically to distinguish the three quarks in a hadron, so the creation operator must be (anti)symmetric under permutations of the three quarks. The physical space is generated by creation operators of the form

$$|f\rangle = |x, q_1, \mu_1; x, q_2, \mu_2; x, q_3, \mu_3\rangle \qquad 19.3$$

where the semi-colons indicates that the state has been symmetrised. We write the creation operator more compactly

$$|f\rangle = |x, i, q_i, \mu_i\rangle = |x, q_1, \mu_1; q_2, \mu_2; q_3, \mu_3\rangle$$

Then the field operator for a baryon is

$$\psi(x) = \psi_{i\mu}(x) = \overline{|x, i, q_i, \mu_i\rangle} + \underline{\langle x, i, q_i, \mu_i|} \qquad 19.4$$

This expression for the field operator is formally like that used when we consider colour compositions of singlet quarks, and to that treated in the literature, e.g. [15](15.19). We draw attention to this because, although formally similar, the symmetry states are interpreted a little differently from the standard model where fields are considered separately from the Fock space on which they act. Historically $SU(2)$, or isospin, was introduced by considering a nucleon as a particle which could be either proton or neutron. But the symmetry in 19.4 was found by considering a particle which is a composition of colours. In other words 19.4 represents the field operator for a particle containing three colours in symmetrical combinations, not one of any colour with symmetrical probability amplitude.

Following the construction of qed we seek to construct an interaction operators $H(x)$ in which a vector boson is emitted or absorbed by the baryon. The treatment of the electromagnetic interaction follows that for leptons, but the Dirac adjoint is

$$\hat{\psi}_{i\mu}(x) = \psi^\dagger_{i\alpha}(x)\gamma^0_{\alpha\mu}$$

and there is a vector current for each quark

$$j^i_\alpha(x) = :\hat{\psi}_{i\mu}(x)\gamma^\alpha_{\mu\nu}\psi_{i\nu}(x):$$

# Discrete Quantum Electrodynamics

Then we can write down the electromagnetic interaction density

$$H(x) = e^i j^i(x) \cdot A(x) \qquad 19.5$$

where $e^i$ is the charge of $q_i$ and we sum over index $i$ so the interaction is colourless. To see that 19.5 reduces to the standard treatment of the electromagnetism we need to establish that the quarks of each colour are decoupled in interactions.

**Theorem:** *In the electromagnetic interaction, in the case when the hadron is preserved (i.e. ignoring pair creation and pair annihilation) the photon interacts with one quark, and leaves spin and momentum of the other quark(s) unchanged.*

*Proof:* It is sufficient to demonstrate the theorem for $i = 1$. The proof for $i = 2$ and 3 is identical. The $i = 1$ term of 19.5 is

$$\sum_{x \in \mathbf{N}} e^i : \hat{\psi}_{i\mu}(x) \gamma^\alpha_{\mu\nu} \psi_{i\nu}(x) : A_\alpha(x) \qquad 19.6$$

Then suppressing $q_i$ in 19.4, 19.6 can be written

$$\sum_{x \in \mathbf{N}} :e_1(|x, \mu_1; \mu_2; \mu_3\rangle + \langle x, \mu_1; \mu_2; \mu_3|)\gamma_{\mu_1\nu} \cdot A(x)(|\overline{x, \nu; \mu_2; \mu_3}\rangle + \langle x, \nu; \mu_2; \mu_3|): \qquad 19.7$$

Ignoring pair creation and annihilation and the antiparticle term, the interaction of a particle is

$$\sum_{x \in \mathbf{N}} :e_1|x, \mu_1; \mu_2; \mu_3\rangle \gamma_{\mu_1\nu} \cdot A(x) \langle x, \nu; \mu_2; \mu_3|:$$

which is in matrix notation

$$\begin{bmatrix} \sum_{x \in \mathbf{N}} e_1 |x, \mu_1\rangle \gamma_{\mu_1\nu'} \gamma^0_{\nu'\nu} \cdot A(x) \langle x, \nu| \\ \sum_{x \in \mathbf{N}} |x, \mu_2\rangle \langle x, \mu_2| \\ \sum_{x \in \mathbf{N}} |x, \mu_3\rangle \langle x, \mu_3| \end{bmatrix} + \text{colour symmetry states} \qquad 19.8$$

Using the resolution of unity and 16.3, which holds for Dirac particles, 19.8 shows that the first quark participates in an interaction with an identical form to the electromagnetic interaction for leptons, while the other two particles are unaffected.

It has been observed that local $U(1)$ transformations have no physical meaning in the description of Hilbert space, so that this must be a symmetry group. Under the $U(1)$ transformation

$$f \to e^{i\alpha(x)} f \qquad 19.9$$

19.1 transforms to

$$e^{i\alpha(x)}(i\partial \cdot \Gamma - \partial \cdot \Gamma \alpha(x) + H(x))f = m e^{i\alpha(x)} f \qquad 19.10$$

since $H(x)$ commutes with $e^{i\alpha(x)}$ which is a function, not an operator. Then we can restore 19.1 by

$$A(x) \to A(x) + \partial \cdot \Gamma \alpha(x)$$

Then replacing $A$ by its expectation and using 17.12 shows that 19.1 is invariant for the classical electromagnetic field.

The electromagnetic interaction, 19.5, preserves colour symmetry because there is an interaction term in each colour index. This is the $U(1)$ subgroup. There is also an $SU(3)$ subgroup for states consisting of three quarks. So we may also try a Yang-Mills interaction in the form of 19.5, but with a current defined for each of the eight generators, $t^a$, of $SU(3)$

$$j_\alpha^a(x) = :\hat{\psi}_{i\mu}(x) t_{ij}^a \gamma_{\mu\nu}^\alpha \psi_{j\nu}(x): \qquad 19.11$$

Because these currents mix flavours the same coupling constant is needed for all quarks, and the interaction density takes the form

$$H(x) = g j^a(x) \cdot A^a(x)$$

where $A^a(x)$ are the gluon fields, which are identical to the photon, up to interactions. Then the equation of motion, 19.1 is

$$(i\partial \cdot \gamma + g j^a(x) \cdot A^a(x)) f = mf \qquad 19.12$$

As with qed the equation of motion, 19.12, can be read as an eigenvalue equation for the gauge covariant derivative showing the origin of gauge invariance. The pure boson vertices are absent from 19.12. Bosonic interactions violate the Ward identity which follows immediately from the equation of motion 9.7 for intermediate vector bosons. Normally the Faddev-Popov ghosts restore the Ward identity but in this model the pure boson vertices are also understood as ghost interactions arising from the quantisation of a gauge field which has no physical meaning.

In the absence of interaction the equation of motion, 19.1, shows that states in the form of 19.2 evolve as three separate Dirac particles; since quarks may have different masses the wave functions cannot in general remain identical. Nonetheless quarks are confined in measured states because there is only one space coordinate in the interaction density for all the quarks in an interaction. If one quark is localised, for example by an electromagnetic interaction with other matter such as might be used in a measurement of position, then the other quarks are confined at the same point when the interaction takes place. Confinement of this sort affects measured states, and should not be seen as a new law of quantum mechanics but rather as an instance of the collapse of the wave function. The interactions which go towards measurement are the same as other interactions, and the projection operator corresponding to the collapse of a wave function is always a composition of currents such as 19.11.

By following through the derivation of Feynman rules (section 18) we observe that the interaction generates a momentum conserving delta function for each individual quark. Thus for hadrons in which the quarks are in eigenstates of momentum, the interaction leaves the momenta and the spin states of two quarks in a baryon unchanged, as though only one of the three quarks participates in the interaction. Thus quarks are `quasi free' – they are confined in co-ordinate space, and yet have values of momentum independent of each other.

### Acknowledgements

I should like to thank a number of physicists who have discussed the content and ideas of this paper on usenet, particularly Paul Colby, Matthew Nobes, Michael Weiss and Toddlius Desiato for their constructive criticism of earlier versions of the paper, John Baez for instruction and advice about the current status of field theory, and the moderators of sci.physics.research (John Baez, Matt McIrvin, Ted Bunn & Philip Helbig) for their vigilance in pointing out lack of clarity in expression in describing the model.